# Photometric mass and mass decomposition in early-type lens galaxies

C. Grillo[1,2,3,4], R. Gobat[3], M. Lombardi[3,4], and P. Rosati[3]

[1] Universitäts-Sternwarte München, Scheinerstr. 1, D-81679 München, Germany
e-mail: cgrillo@usm.lmu.de
[2] Max-Planck-Institut für extraterrestrische Physik, Giessenbachstr., D-85748, Garching bei München, Germany

[3] European Southern Observatory, Karl-Schwarzschild-Str. 2, D-85748, Garching bei München, Germany

[4] Università degli Studi di Milano, Department of Physics, via Celoria 16, I-20133 Milan, Italy

**ABSTRACT**

*Context.* The largest spectroscopically selected sample of strong gravitational lens systems presented and analyzed to date is that of the Sloan Lens ACS (SLACS) survey. For the 57 massive early-type lens galaxies in the sample, photometric and spectroscopic measurements are available from the Sloan Digital Sky Survey (SDSS).
*Aims.* By using the SDSS multicolor photometry and lens modeling, we study stellar-mass properties and the luminous and dark matter composition of the early-type lens galaxies in the sample.
*Methods.* We fit the lens spectral energy distributions (SEDs) consisting of *ugriz* magnitudes with a three-parameter grid (age, star-formation timescale, and photometric mass) of Bruzual & Charlot's and Maraston's composite stellar-population models, computed by adopting solar metallicity and various initial mass functions (IMFs). We also utilize the best-fit parameters derived from the lens models of the total projected mass enclosed within the disk defined by the Einstein radius of each system.
*Results.* We observe that early-type lens galaxies have the same physical properties as massive non-lens galaxies. In particular, we investigate the relationships between stellar mass and both the size and the surface stellar-mass density of the lens galaxies, which are consistent with those determined for non-lens galaxies in the local Universe. We find statistically significant evidence that more luminous and massive lens galaxies tend to form in regions of higher galaxy density, as for all early-type galaxies. Interestingly, for the corresponding stellar quantities we measure the same scaling law between effective mass-to-light ratio and mass used to explain the "tilt" of the Fundamental Plane (FP), and the same evolution in the effective mass-to-light ratio with redshift as derived from the FP. We conclude that the total (luminous+dark) mass of the lenses is linearly proportional to the luminous mass, at a confidence level of more than 99%. In addition, by assuming that the lens galaxies are homologous systems, we study their distribution of dark matter and estimate a value on the order of 30% for the dark over total projected mass fraction contained within the average Einstein ring of radius of approximately 4 kpc.
*Conclusions.* These results suggest that early-type lens galaxies are an unbiased subsample representative of the complete sample of early-type galaxies. This allows us to generalize our findings on the amount and distribution of dark matter in lens galaxies to the population of massive early-type galaxies. For the assumed metallicity, we note that a Salpeter IMF is better suited than either a Chabrier or Kroupa IMF to describing the sample of lenses.

## 1. Introduction

Despite their fairly simple morphological appearance, early-type galaxies are complex astrophysical objects. In addition to their luminous stellar component, significative evidence of a dark matter component (in the form of a diffuse halo) exists, especially in the outer regions of the most massive systems (e.g., Treu & Koopmans 2004; Gavazzi et al. 2007; Grillo et al. 2008c). Although baryons are estimated to represent only a small fraction of the total mass in these galaxies, it is now commonly agreed that baryons play a crucial role in the formation and evolution of dark matter haloes. For instance, the accretion of dissipational gas in the inner regions ($\sim 10$ kpc) of massive early-type galaxies has been recognized to be the possible origin of a dark-matter density profile that appears steeper (e.g., Jesseit et al. 2002; Gnedin et al. 2004; Kazantzidis et al. 2004) than predicted by dark-matter-only cosmological simulations

*Send offprint requests to*: C. Grillo

(Navarro et al. 1997). The time required to convert gas into collisionless stars is also an essential parameter for the evolution of these systems within the hierarchical formation scenario (e.g., van Dokkum et al. 1999).

Notwithstanding, the distribution of luminous and dark matter in early-type galaxies exhibits a strong degree of homogeneity (e.g., Treu et al. 2006; Grillo et al. 2008b; Bolton et al. 2008b) such that several empirical scaling laws between photometric and spectroscopic properties can remain valid (e.g., Faber & Jackson 1976; Kormendy 1977; Dressler et al. 1987; Baum 1959; Bower et al. 1992; Bender et al. 1992, 1993; Ferrarese & Merritt 2000; Gebhardt 2000; Shen et al. 2003; McIntosh et al. 2005). Among these scaling laws, an extensively studied relation between effective radius, central stellar velocity dispersion, and average surface brightness is known as the Fundamental Plane (FP; Djorgovski & Davies 1987; Dressler et al. 1987). The interplay between luminous and dark matter remains to be investigated in depth to fully understand the origin of the observed scaling laws and, ultimately, the mechanisms responsible for the initial formation and the subsequent evolution of early-type galaxies.

In the field environment, mass seems to be the main parameter driving the star formation activity (e.g., Cowie et al. 1996) and, in general, galaxy formation and evolution, while environment probably plays only a secondary role (e.g., Moran et al. 2005; Yee et al. 2005). In contrast, in clusters the stellar population properties of early-type galaxies also depend significantly on the environment (e.g., Gobat et al. 2008, Rettura et al. 2008). Measurements of the luminous mass (i.e., in the form of stars) of early-type galaxies can be performed by fitting their observed SEDs with composite stellar-population models (e.g., Fontana et al. 2004; Rocca-Volmerange et al. 2004; Saracco et al. 2004). However, the most obvious source of information on the total (i.e., luminous+dark) mass of local early-type galaxies is provided by different dynamical tracers, such as stars (e.g., Saglia et al. 1992; Gerhard et al. 2001), cold atomic hydrogen and warm ionized gas (e.g., Buson et al. 1993; Franx et al. 1994), X-ray emitting gas (e.g., Mushotzky et al. 1994; Loewenstein & White 1999), globular clusters (e.g., Mould et al. 1990), and planetary nebulae (e.g., Arnaboldi et al. 1998). Furthermore, in early-type galaxies acting as lenses, strong and weak gravitational lensing analyses (e.g., Langston et al. 1990; Rix et al. 1992; Gavazzi et al. 2007; Grillo et al. 2008c) offer the most accurate projected total mass measurements in extragalactic astronomy. By combining (e.g., Trott & Webster 2002; Treu & Koopmans 2004; Koopmans et al. 2006) or by comparing (e.g., Drory et al. 2004; Ferreras et al. 2005; Rettura et al. 2006) these various mass diagnostics, it is possible to study in detail the internal structure of early-type galaxies and to test galaxy formation and evolution models (e.g., Nagamine et al. 2004; De Lucia et al. 2006).

The paper is organized as follows. In Sect. 2, we describe the sample of early-type lens galaxies that are considered in our study. Then, in Sect. 3, we investigate some physical properties of the lenses, using different composite stellar-population models to fit the SDSS multi-band photometry. In Sect. 4, by adding lensing information, we address the mass decomposition in terms of luminous and dark matter contained within the Einstein radii of the galaxies in the sample. Finally, in Sect. 5 we summarize the results obtained in this paper. Throughout this work, we assume a concordance cosmology with Hubble parameter $H_0 = 70$ km s$^{-1}$ Mpc$^{-1}$, and matter and dark-energy density parameters $\Omega_\mathrm{m} = 0.3$, $\Omega_\Lambda = 0.7$. In this study, all logarithms have base 10 and adimensional arguments, which are obtained, where necessary, by dividing the dimensional quantities by the adopted measurement units.

## 2. The sample of early-type lens galaxies

In this analysis, we focus on the "grade-A" strong gravitational lensing systems discovered in the SLACS survey[1] and presented in Bolton et al. (2008a). These systems were first spectroscopically selected from the database of the SDSS[2] by identifying emission lines of a hypothetical source located at a redshift higher than that of the possible lens galaxy responsible for the continuum. Then, they were all observed at least once through the F814W filter of the Advanced Camera for Surveys (ACS) onboard the *Hubble Space Telescope* (HST), to confirm the lens hypothesis. For more details of the selection procedure, observations, and lens modeling, we refee the reader to Bolton et al. (2006, 2008a).

We restrict our sample to the 57 massive early-type galaxies acting as lenses. These galaxies belong to the SDSS Main galaxy sample (Strauss et al. 2002) and luminous red galaxy (LRG) sample (Eisenstein et al. 2001). In Table 1, we show the spectroscopic and photometric quantities relevant to this work of the galaxies of the sample where we note that the effective angle $\theta_\mathrm{e}$ is measured by fitting a de Vaucouleurs luminosity profile on the F814W images. The foreground (lens) galaxies have redshifts $z_l$ between 0.06 and 0.51; the background (source) objects have redshifts $z_s$ between 0.20 and 1.19. The corresponding angular diameter distances are labeled as $D_\mathrm{ol}$ and $D_\mathrm{os}$, respectively. The values of the Einstein angle $\theta_\mathrm{Ein}$ of the lensing systems range from $0.69''$ to $1.78''$ ($\langle R_\mathrm{Ein} \rangle = \langle D_\mathrm{ol}\, \theta_\mathrm{Ein} \rangle = 4.1 \pm 0.2$ kpc). The values of the Einstein angle are typically 0.6 times the values of the effective angle $\theta_\mathrm{e}$ of the lens galaxies. The lenses have *ugriz* SDSS photometry and measurements of the central stellar velocity dispersion (defined as the luminosity-weighted average velocity dispersion of stars inside an aperture of radius $R_\mathrm{e}/8$, $\langle \sigma_0 \rangle = 271 \pm 7$ km s$^{-1}$) from the SDSS. They have been shown to be luminous and massive galaxies, similar to early-type non-lens galaxies in terms of redshift, stellar velocity dispersion, total mass density profile, and environment (see Treu et al. 2006, 2008; Koopmans et al. 2006; Bolton et al. 2008a, 2008b).

## 3. Stellar population-synthesis models

To investigate properties related to stellar populations and measure the photometric (stellar) masses of the sample galaxies, we compare their observed SEDs, derived from the SDSS data, with composite stellar-population models computed using two different sets of solar-metallicity templates.

[1] http://www.slacs.org/

[2] http://www.sdss.org/

**Table 1.** The relevant spectroscopic and photometric measurements of the 57 early-type grade-A lens galaxies of the SLACS survey.

| SDSS Name | $z_l$ | $z_s$ | $\theta_{\rm Ein}$ ($''$) | $\theta_{\rm e}$ ($''$) | $u$ (mag) | $g$ (mag) | $r$ (mag) | $i$ (mag) | $z$ (mag) |
|---|---|---|---|---|---|---|---|---|---|
| J0008−0004 | 0.440 | 1.192 | 1.16 | 1.71 | 22.632 ± 1.095 | 20.655 ± 0.071 | 19.294 ± 0.030 | 18.654 ± 0.027 | 18.082 ± 0.060 |
| J0029−0055 | 0.227 | 0.931 | 0.96 | 2.16 | 20.479 ± 0.163 | 18.792 ± 0.019 | 17.546 ± 0.010 | 17.072 ± 0.010 | 16.728 ± 0.022 |
| J0037−0942 | 0.196 | 0.632 | 1.53 | 2.19 | 19.780 ± 0.127 | 18.038 ± 0.010 | 16.807 ± 0.006 | 16.340 ± 0.006 | 15.993 ± 0.016 |
| J0044+0113 | 0.120 | 0.197 | 0.79 | 2.61 | 18.701 ± 0.043 | 17.120 ± 0.005 | 16.195 ± 0.004 | 15.771 ± 0.004 | 15.461 ± 0.008 |
| J0109+1500 | 0.294 | 0.525 | 0.69 | 1.38 | 22.867 ± 1.225 | 19.753 ± 0.032 | 18.204 ± 0.013 | 17.641 ± 0.012 | 17.261 ± 0.036 |
| J0157−0056 | 0.513 | 0.924 | 0.79 | 1.06 | 22.722 ± 0.763 | 21.258 ± 0.105 | 19.670 ± 0.030 | 18.702 ± 0.020 | 18.280 ± 0.043 |
| J0216−0813 | 0.332 | 0.523 | 1.16 | 2.67 | 21.075 ± 0.335 | 19.124 ± 0.024 | 17.456 ± 0.009 | 16.860 ± 0.009 | 16.573 ± 0.021 |
| J0252+0039 | 0.280 | 0.982 | 1.04 | 1.39 | 21.331 ± 0.287 | 20.061 ± 0.033 | 18.813 ± 0.016 | 18.237 ± 0.014 | 17.916 ± 0.030 |
| J0330−0020 | 0.351 | 1.071 | 1.10 | 1.20 | 20.714 ± 0.218 | 19.947 ± 0.039 | 18.462 ± 0.015 | 17.919 ± 0.015 | 17.564 ± 0.031 |
| J0405−0455 | 0.075 | 0.810 | 0.80 | 1.36 | 19.504 ± 0.081 | 17.618 ± 0.008 | 16.771 ± 0.005 | 16.365 ± 0.006 | 16.060 ± 0.011 |
| J0728+3835 | 0.206 | 0.688 | 1.25 | 1.78 | 20.404 ± 0.135 | 18.623 ± 0.014 | 17.356 ± 0.007 | 16.887 ± 0.007 | 16.583 ± 0.017 |
| J0737+3216 | 0.322 | 0.581 | 1.00 | 2.82 | 21.239 ± 0.266 | 19.400 ± 0.025 | 17.834 ± 0.010 | 17.214 ± 0.008 | 16.872 ± 0.020 |
| J0822+2652 | 0.241 | 0.594 | 1.17 | 1.82 | 20.548 ± 0.147 | 18.899 ± 0.016 | 17.501 ± 0.007 | 16.986 ± 0.007 | 16.627 ± 0.015 |
| J0903+4116 | 0.430 | 1.065 | 1.29 | 1.78 | 21.401 ± 0.244 | 20.302 ± 0.041 | 18.646 ± 0.016 | 17.967 ± 0.015 | 17.540 ± 0.031 |
| J0912+0029 | 0.164 | 0.324 | 1.63 | 3.87 | 19.327 ± 0.063 | 17.410 ± 0.007 | 16.229 ± 0.004 | 15.746 ± 0.004 | 15.379 ± 0.008 |
| J0935−0003 | 0.347 | 0.467 | 0.87 | 4.24 | 21.331 ± 0.399 | 19.193 ± 0.028 | 17.515 ± 0.010 | 16.915 ± 0.009 | 16.554 ± 0.022 |
| J0936+0913 | 0.190 | 0.588 | 1.09 | 2.11 | 20.073 ± 0.120 | 18.218 ± 0.010 | 17.002 ± 0.006 | 16.555 ± 0.006 | 16.226 ± 0.015 |
| J0946+1006 | 0.222 | 0.609 | 1.38 | 2.35 | 20.465 ± 0.171 | 18.899 ± 0.016 | 17.578 ± 0.008 | 17.102 ± 0.008 | 16.744 ± 0.021 |
| J0956+5100 | 0.240 | 0.470 | 1.33 | 2.19 | 20.175 ± 0.136 | 18.475 ± 0.012 | 17.129 ± 0.007 | 16.633 ± 0.006 | 16.247 ± 0.013 |
| J0959+4416 | 0.237 | 0.531 | 0.96 | 1.98 | 20.452 ± 0.150 | 18.850 ± 0.014 | 17.510 ± 0.008 | 17.016 ± 0.007 | 16.698 ± 0.020 |
| J0959+0410 | 0.126 | 0.535 | 0.99 | 1.39 | 20.403 ± 0.088 | 18.697 ± 0.012 | 17.639 ± 0.007 | 17.168 ± 0.006 | 16.763 ± 0.016 |
| J1016+3859 | 0.168 | 0.439 | 1.09 | 1.46 | 20.388 ± 0.095 | 18.434 ± 0.010 | 17.265 ± 0.006 | 16.827 ± 0.006 | 16.502 ± 0.013 |
| J1020+1122 | 0.282 | 0.553 | 1.20 | 1.59 | 21.470 ± 0.239 | 19.508 ± 0.020 | 18.003 ± 0.010 | 17.465 ± 0.009 | 17.172 ± 0.021 |
| J1023+4230 | 0.191 | 0.696 | 1.41 | 1.77 | 20.390 ± 0.103 | 18.664 ± 0.013 | 17.366 ± 0.007 | 16.871 ± 0.007 | 16.548 ± 0.016 |
| J1029+0420 | 0.104 | 0.615 | 1.01 | 1.56 | 19.340 ± 0.048 | 17.556 ± 0.006 | 16.629 ± 0.004 | 16.224 ± 0.004 | 15.884 ± 0.008 |
| J1100+5329 | 0.317 | 0.858 | 1.52 | 2.24 | 21.003 ± 0.311 | 19.161 ± 0.021 | 17.704 ± 0.010 | 17.103 ± 0.009 | 16.869 ± 0.029 |
| J1106+5228 | 0.096 | 0.407 | 1.23 | 1.68 | 18.841 ± 0.034 | 16.940 ± 0.005 | 16.001 ± 0.003 | 15.612 ± 0.003 | 15.287 ± 0.006 |
| J1112+0826 | 0.273 | 0.629 | 1.49 | 1.50 | 21.775 ± 0.377 | 19.375 ± 0.025 | 17.809 ± 0.009 | 17.243 ± 0.008 | 16.906 ± 0.016 |
| J1134+6027 | 0.153 | 0.474 | 1.10 | 2.02 | 19.936 ± 0.089 | 18.085 ± 0.008 | 17.024 ± 0.005 | 16.538 ± 0.005 | 16.192 ± 0.010 |
| J1142+1001 | 0.222 | 0.504 | 0.98 | 1.91 | 20.711 ± 0.231 | 18.836 ± 0.017 | 17.495 ± 0.008 | 17.002 ± 0.007 | 16.723 ± 0.020 |
| J1143−0144 | 0.106 | 0.402 | 1.68 | 4.80 | 18.635 ± 0.032 | 16.845 ± 0.005 | 15.827 ± 0.003 | 15.405 ± 0.003 | 15.062 ± 0.006 |
| J1153+4612 | 0.180 | 0.875 | 1.05 | 1.16 | 20.575 ± 0.110 | 18.793 ± 0.015 | 17.680 ± 0.008 | 17.218 ± 0.008 | 16.895 ± 0.015 |
| J1204+0358 | 0.164 | 0.631 | 1.31 | 1.47 | 20.330 ± 0.093 | 18.544 ± 0.012 | 17.399 ± 0.006 | 16.936 ± 0.006 | 16.661 ± 0.012 |
| J1205+4910 | 0.215 | 0.481 | 1.22 | 2.59 | 20.744 ± 0.250 | 18.541 ± 0.012 | 17.234 ± 0.007 | 16.718 ± 0.006 | 16.343 ± 0.013 |
| J1213+6708 | 0.123 | 0.640 | 1.42 | 3.23 | 19.054 ± 0.052 | 17.176 ± 0.006 | 16.159 ± 0.004 | 15.731 ± 0.004 | 15.377 ± 0.008 |
| J1218+0830 | 0.135 | 0.717 | 1.45 | 3.18 | 19.223 ± 0.066 | 17.394 ± 0.006 | 16.343 ± 0.004 | 15.893 ± 0.004 | 15.543 ± 0.008 |
| J1250+0523 | 0.232 | 0.795 | 1.13 | 1.81 | 19.983 ± 0.084 | 18.500 ± 0.012 | 17.256 ± 0.007 | 16.733 ± 0.006 | 16.464 ± 0.014 |
| J1402+6321 | 0.205 | 0.481 | 1.35 | 2.70 | 20.393 ± 0.142 | 18.293 ± 0.011 | 16.952 ± 0.006 | 16.444 ± 0.005 | 16.077 ± 0.011 |
| J1403+0006 | 0.189 | 0.473 | 0.83 | 1.46 | 20.371 ± 0.107 | 18.687 ± 0.014 | 17.506 ± 0.007 | 17.028 ± 0.007 | 16.721 ± 0.018 |
| J1416+5136 | 0.299 | 0.811 | 1.37 | 1.43 | 21.225 ± 0.298 | 19.592 ± 0.024 | 18.080 ± 0.011 | 17.521 ± 0.009 | 17.171 ± 0.027 |
| J1420+6019 | 0.063 | 0.535 | 1.04 | 2.06 | 18.196 ± 0.025 | 16.386 ± 0.004 | 15.541 ± 0.003 | 15.153 ± 0.003 | 14.869 ± 0.016 |
| J1430+4105 | 0.285 | 0.575 | 1.52 | 2.55 | 20.331 ± 0.104 | 18.974 ± 0.012 | 17.706 ± 0.007 | 17.196 ± 0.006 | 16.851 ± 0.014 |
| J1436−0000 | 0.285 | 0.805 | 1.12 | 2.24 | 21.250 ± 0.267 | 19.216 ± 0.020 | 17.801 ± 0.009 | 17.229 ± 0.009 | 16.876 ± 0.023 |
| J1443+0304 | 0.134 | 0.419 | 0.81 | 0.94 | 20.607 ± 0.122 | 18.524 ± 0.011 | 17.483 ± 0.006 | 17.034 ± 0.006 | 16.696 ± 0.013 |
| J1451−0239 | 0.125 | 0.520 | 1.04 | 2.48 | 19.240 ± 0.091 | 17.538 ± 0.008 | 16.480 ± 0.005 | 16.129 ± 0.004 | 15.673 ± 0.009 |
| J1525+3327 | 0.358 | 0.717 | 1.31 | 2.90 | 20.936 ± 0.220 | 19.446 ± 0.022 | 17.877 ± 0.009 | 17.247 ± 0.008 | 16.935 ± 0.019 |
| J1531−0105 | 0.160 | 0.744 | 1.71 | 2.50 | 19.475 ± 0.122 | 17.519 ± 0.009 | 16.362 ± 0.005 | 15.915 ± 0.005 | 15.591 ± 0.009 |
| J1538+5817 | 0.143 | 0.531 | 1.00 | 1.58 | 19.495 ± 0.059 | 18.174 ± 0.009 | 17.173 ± 0.006 | 16.741 ± 0.006 | 16.429 ± 0.012 |
| J1621+3931 | 0.245 | 0.602 | 1.29 | 2.14 | 20.502 ± 0.226 | 18.780 ± 0.017 | 17.380 ± 0.010 | 16.859 ± 0.010 | 16.553 ± 0.019 |
| J1627−0053 | 0.208 | 0.524 | 1.23 | 1.98 | 20.623 ± 0.190 | 18.588 ± 0.017 | 17.286 ± 0.008 | 16.805 ± 0.008 | 16.490 ± 0.017 |
| J1630+4520 | 0.248 | 0.793 | 1.78 | 1.96 | 20.594 ± 0.138 | 18.877 ± 0.015 | 17.396 ± 0.007 | 16.862 ± 0.007 | 16.541 ± 0.014 |
| J1636+4707 | 0.228 | 0.674 | 1.09 | 1.68 | 20.980 ± 0.187 | 19.005 ± 0.013 | 17.665 ± 0.007 | 17.174 ± 0.007 | 16.877 ± 0.022 |
| J2238−0754 | 0.137 | 0.713 | 1.27 | 2.33 | 19.716 ± 0.116 | 17.803 ± 0.009 | 16.766 ± 0.005 | 16.309 ± 0.005 | 15.963 ± 0.011 |
| J2300+0022 | 0.229 | 0.464 | 1.24 | 1.83 | 20.517 ± 0.190 | 19.007 ± 0.017 | 17.647 ± 0.009 | 17.127 ± 0.008 | 16.783 ± 0.022 |
| J2303+1422 | 0.155 | 0.517 | 1.62 | 3.28 | 19.467 ± 0.194 | 17.562 ± 0.012 | 16.385 ± 0.006 | 15.907 ± 0.006 | 15.585 ± 0.014 |
| J2321−0939 | 0.082 | 0.532 | 1.60 | 4.11 | 18.085 ± 0.037 | 16.145 ± 0.004 | 15.200 ± 0.003 | 14.771 ± 0.003 | 14.458 ± 0.006 |
| J2341+0000 | 0.186 | 0.807 | 1.44 | 3.15 | 19.513 ± 0.119 | 18.121 ± 0.013 | 16.921 ± 0.007 | 16.381 ± 0.007 | 16.010 ± 0.018 |

Notes – Magnitudes are extinction-corrected modelMag (AB) from the SDSS.
References – Bolton et al. (2008a).

We use the multi-band photometry ($u$, $g$, $r$, $i$, and $z$ bands, extending from 354 to 913 nm) of the SDSS. The extinction-corrected total magnitudes, called modelMag in the SDSS catalog, are obtained by first fitting a de Vaucouleurs profile,

$$I(R) = I_0 \exp\left[-7.67 \left(\frac{R}{R_{\rm e}}\right)^{1/4}\right] , \qquad (1)$$

($R$ and $R_{\rm e} = D_{\rm ol}\,\theta_{\rm e}$ being, respectively, the two-dimensional projected radial coordinate and the standard optical effective radius) to the $r$-band luminosity profile of each galaxy and then varying this model, only in amplitude, for the other bands (after convolution with the relevant point spread function). The resulting magnitudes (see Table 1) correspond to magnitudes measured through equivalent apertures in all the bands and thus provide unbiased galaxy colors, in the absence of color gradients. Since we study the average luminosity-weighted stellar population of each galaxy, a possible color gradient would only produce slightly higher photometric errors and thus have a negligible impact on the photometric mass estimates determined in the following. We have checked that the measurements of the magnitudes are not significantly contaminated by the lensed objects. The lensed images are indeed two or more magnitudes fainter than the lens galaxies. In addition, since the estimates of the magnitudes are performed by fitting a de Vaucouleurs profile, this reduces the possible effect of contamination.

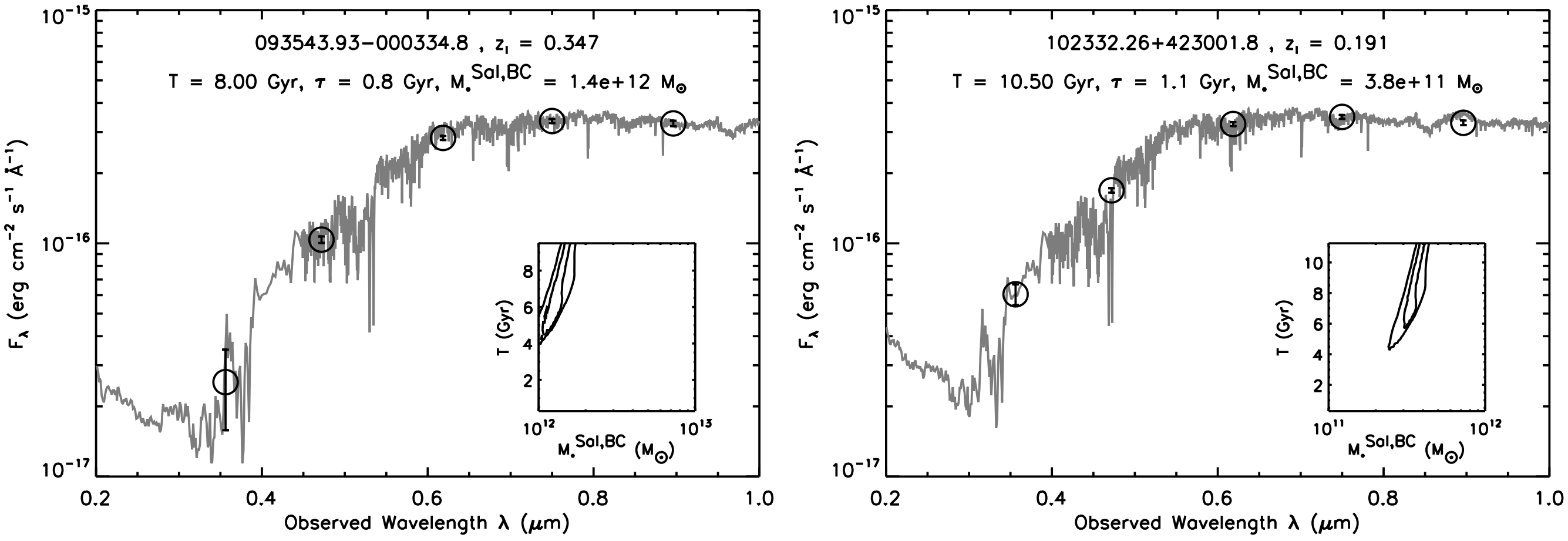

**Fig. 1.** SEDs and best-fit models of the two lens galaxies SDSS J0935−0003 (*left*) and SDSS J1023+4230 (*right*). The observed total flux densities and their 1 $\sigma$ errors, as measured by the SDSS ($u$, $g$, $r$, $i$, and $z$ filters), are represented by circles and error bars. In each panel, we show, at the top, the complete coordinates and the best-fit values of the parameters of the grid ($T, \tau, M_*$), and at the bottom, the 68% and 99% confidence regions on the plane $T - M_*$.

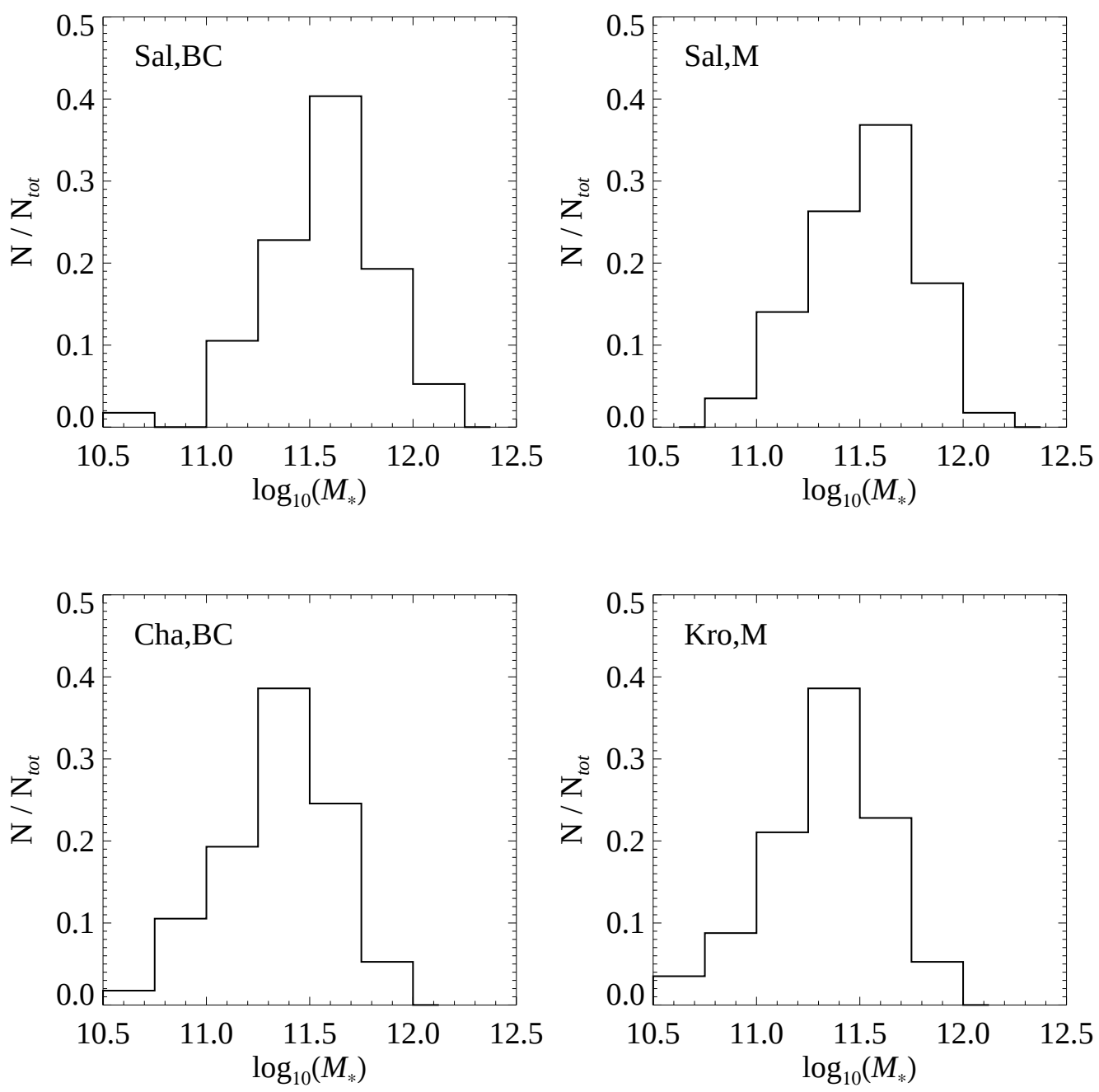

**Fig. 2.** Histograms of the best-fit values of the photometric mass derived by comparing the observed SEDs of the 57 early-type grade-A lens galaxies of the SLACS survey with different composite stellar-population models [Bruzual & Charlot (BC) and Maraston (M)] and IMFs [Salpeter (Sal), Chabrier (Cha), Kroupa (Kro)].

To compute our composite stellar-population models, we choose Bruzual & Charlot (2003; indexed BC) and Maraston (2005; indexed M) dust-free templates at solar metallicity. We adopt a Salpeter (1955; indexed Sal) and a Chabrier (2003; indexed Cha) IMF for Bruzual & Charlot models and a Salpeter and a Kroupa (2001; indexed Kro) IMF for Maraston models. These IMFs differ significantly only for stars less massive than 1 $M_\odot$. The lower and upper limits to the IMFs are usually fixed by observational constraints and considerations of stellar structure to 0.1 and 100 $M_\odot$, respectively. We assume the standard parametrization of the SFH called a delayed exponential given by

$$\psi(t) = \frac{t}{\tau^2} \exp\left(-\frac{t}{\tau}\right) , \qquad (2)$$

where $t$ is a time variable that is positive from the onset of star formation and increases towards lower redshift ($t$ must not be confused with the cosmic time) and $\tau$ is the characteristic timescale of the star formation. This SFH, which is more realistic than a simple exponentially declining one (Gavazzi et al. 2002), is similar to that proposed by Sandage (1986). In population-synthesis studies, it is possible to convert the star-formation-weighted age, defined as

$$t_f = \frac{\int_0^T \mathrm{d}t' \, (T - t') \, \psi(t')}{\int_0^T \mathrm{d}t' \, \psi(t')} , \qquad (3)$$

where $T$ is the "age" of the model, i.e., the time elapsed since the onset of star formation (constrained by the age of the Universe at the galaxy redshift), into a formation redshift $z_f$ and to use this to account for the average age of the bulk of the stars in a galaxy. In conclusion, the fit to the observed SEDs is performed with a grid of composite stellar-population models characterized by the three parameters $T$, $\tau$, and photometric mass $M_*$. We measure the uncertainties in the best-fit parameters by projecting the joint probability density distribution onto the corresponding axes. Here we concentrate on the galaxy properties related to the photometric mass and leave the study of the star-formation history to future work. From the best-fit models and by assuming passive evolution, the photometric mass and the $B$-band luminosity at redshift $z = 0$ ($M_{*,0}$ and $L_{B,0}$, respectively) are estimated. These quantities are shown to be important in the following analysis. In Fig. 1, we plot the observed SEDs and the best-fit models of two of the galaxies in the sample. We remark that in the SDSS photometry the lack of possible significant systematical errors caused by calibration or aperture corrections ensure accurate estimates of the best-fit model parameters.

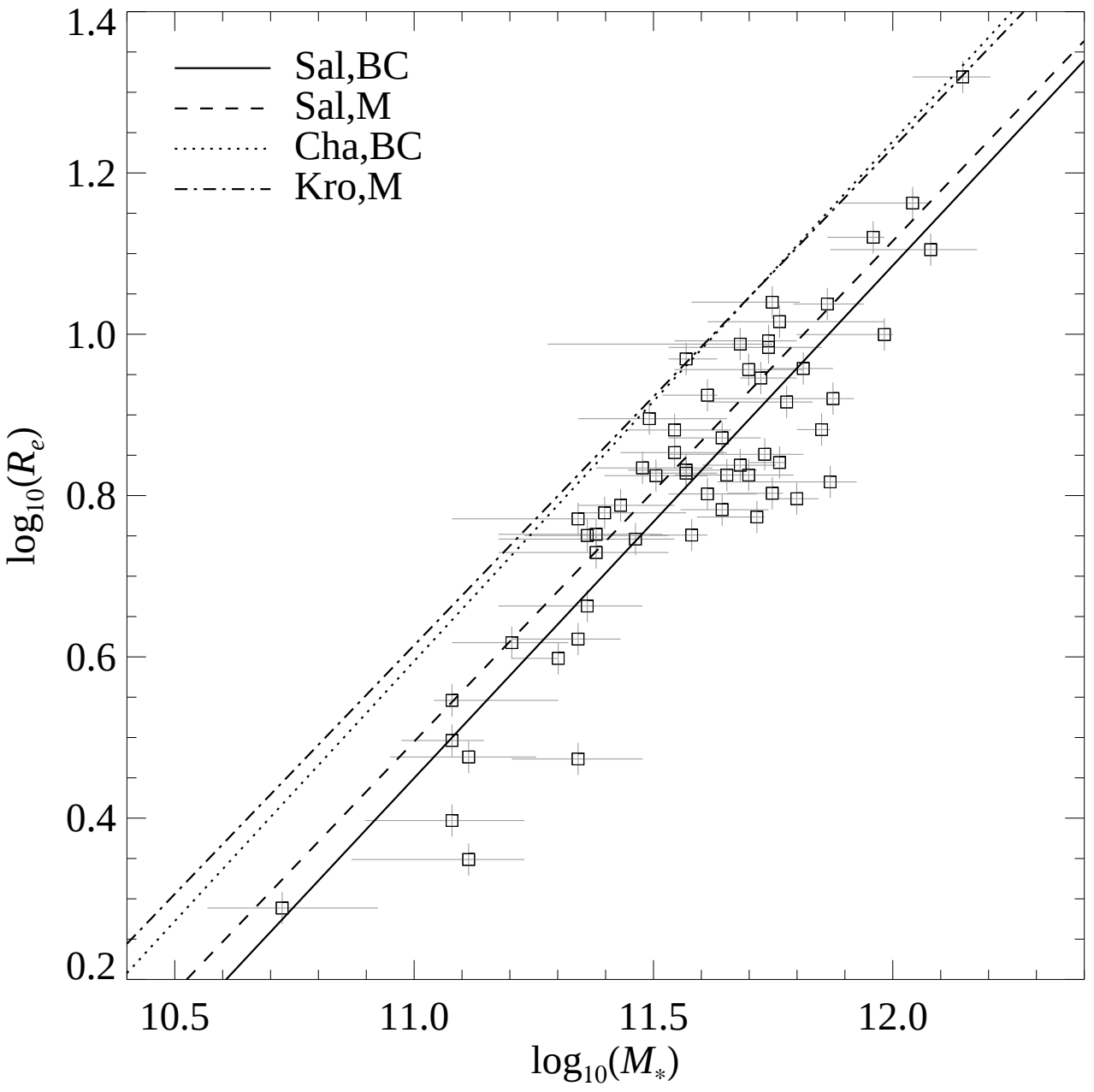

**Fig. 3.** The effective radius versus the photometric mass of the 57 early-type grade-A lens galaxies of the SLACS survey. The points with 1 $\sigma$ error bars, obtained by using Bruzual & Charlot models and a Salpeter IMF, and the best-fit correlation lines for all the different composite stellar-population models [Bruzual & Charlot (BC) and Maraston (M)] and IMFs [Salpeter (Sal), Chabrier (Cha), Kroupa (Kro)] are shown.

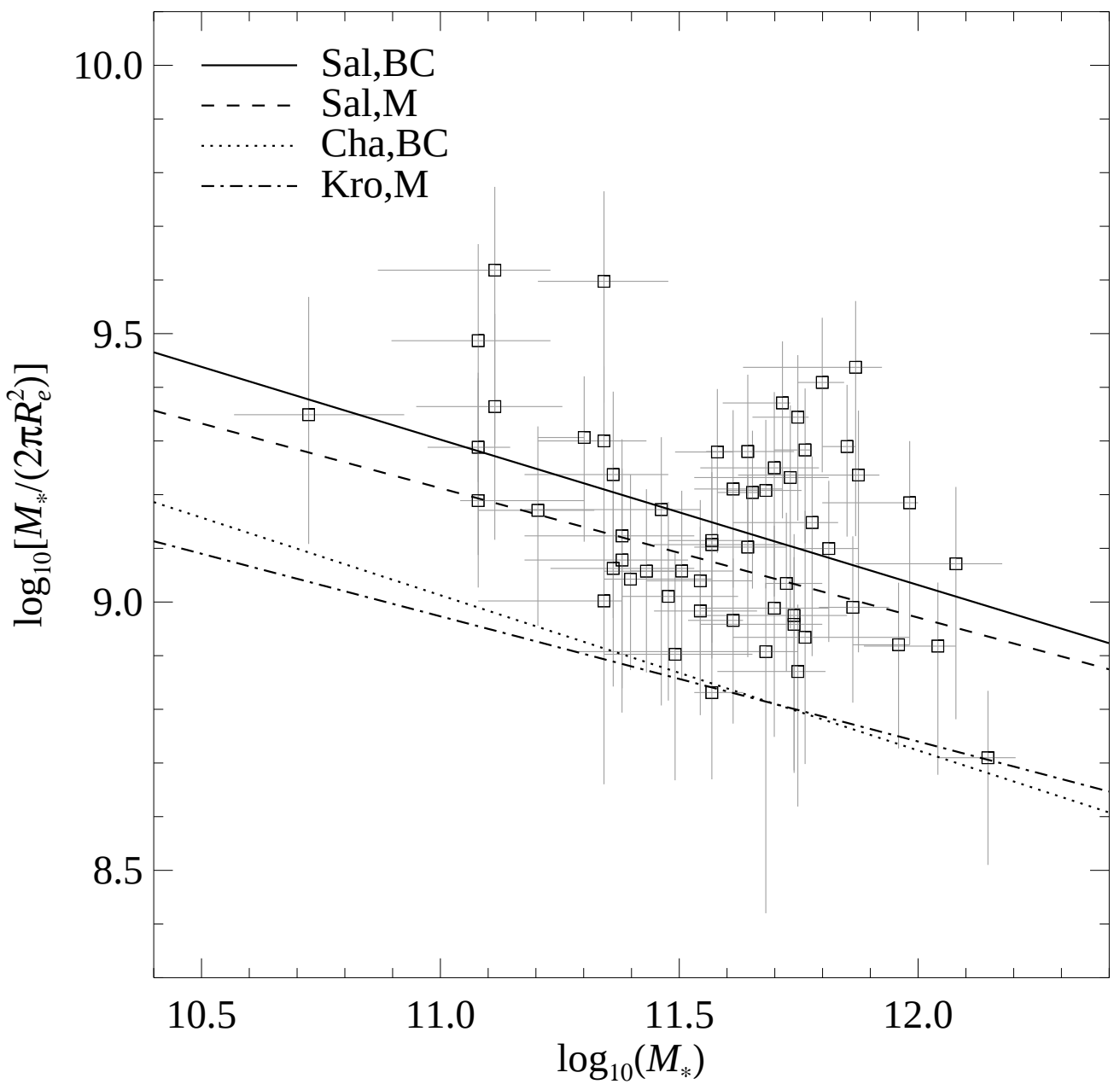

**Fig. 4.** The surface stellar-mass density versus the photometric mass of the 57 early-type grade-A lens galaxies of the SLACS survey. The points with 1 $\sigma$ error bars, obtained by using Bruzual & Charlot models and a Salpeter IMF, and the best-fit correlation lines for all the different composite stellar-population models [Bruzual & Charlot (BC) and Maraston (M)] and IMFs [Salpeter (Sal), Chabrier (Cha), Kroupa (Kro)] are shown.

The best-fit values of the photometric mass for all the galaxies in the sample, obtained by using the different composite stellar-population models and IMFs ($M_*^{\rm Sal,BC}$, $M_*^{\rm Sal,M}$, $M_*^{\rm Cha,BC}$, and $M_*^{\rm Kro,M}$), are reported in Table 4 and plotted in Fig. 2. The measured photometric masses (between $3 \times 10^{10}\, M_\odot$ and $1.4 \times 10^{12}\, M_\odot$, with typical errors on the order of 30%) show that the sample is representative of massive (and luminous) early-type galaxies. In Tables 2 and 4, we see that the photometric mass values depend significantly on the adopted IMF, but not on the stellar population model. In particular, assuming a Salpeter IMF, we find estimates of photometric mass that are more than one and a half times higher than those obtained by choosing either a Chabrier or Kroupa IMF; however, for a Salpeter IMF, Bruzual & Charlot's templates infer photometric mass measurements that are consistent, within the errors, with those derived from Maraston's templates. This last result is not surprising, since the lens galaxies are older than the age range across which the two sets of templates differ.

The first differences are caused by the Chabrier and Kroupa IMFs containing a higher fractional number of massive stars than the Salpeter IMF. At a fixed age and metallicity, this implies that the integrated flux predicted by models with the first two IMFs is higher than the integrated flux predicted by models with a Salpeter IMF. Thus, since the photometric mass is almost a proportionality factor with which we multiply the model flux to reproduce the observed flux of a galaxy, the models with a Chabrier and Kroupa IMF yield lower mass values than the models with a Salpeter IMF.

We emphasize that our photometric mass measurements are not affected in any significant way by the lensed objects. Since the lensed images are bluer than the lens galaxies, the lensed images indeed influence very little the photometry in the redder filters, which are known to be more sensitive to photometric mass estimates. Moreover, we find that the values of the photometric mass obtained by using the complete multi-band photometry or excluding the $u$ magnitudes are approximately the same and consistent within the errors.

On the other hand, we note that the photometric mass estimates depend on the adopted metallicity and, as already discussed, on the IMF. The effects of metallicity and the fraction of massive ($M > 1 M_\odot$) stars on the photometric mass measurements have opposite senses. Indeed, for a given color, low-metallicity models have lower near-infrared fluxes than high-metallicity ones, while models with a top-heavy IMF are brighter than models with a Salpeter one, thus driving the stellar mass estimates up and down, respectively. A metallicity of half or less than solar would thus be required to reproduce, with either a Chabrier or Kroupa IMF, the photometric mass values derived from solar-metallicity models with a Salpeter IMF, which were shown to be consistent with independent luminous mass estimates obtained from a joint strong lensing and stellar dynamics analysis (Grillo et al. 2008a). However, such a low metallicity would be inconsistent with studies of massive ellipticals in the nearby Universe (e.g., Thomas et al. 2005; Gallazzi et al. 2006). By using stellar population models at solar and super-solar metallicity, van Dokkum (2008) inferred that early-type galaxies may instead have a "bottom-light" IMF (deficient in low-mass stars but with different slope than a Chabrier or Kroupa IMF). For old galaxy ages, this type of IMF implies mass-to-light ratios that are consistent with those measured with a Salpeter IMF. Finally, we notice that photometric mass estimates obtained by using single burst or truncated star-formation histories are consistent, given the errors, with those derived from delayed, exponentially declining star formation histories. This is not surprising since our parameter grid also accommodates SSP-like models.

**Table 2.** Important quantities derived from different composite stellar-population models and IMFs.

| | Sal,BC | Sal,M | Cha,BC | Kro,M |
|---|---|---|---|---|
| $\langle M_{*} \rangle$ ($10^{10}$ $M_{\odot}$) | 46 ± 4 | 39 ± 3 | 27 ± 2 | 25 ± 2 |
| $\alpha$ | 0.64 ± 0.04 | 0.62 ± 0.05 | 0.64 ± 0.04 | 0.62 ± 0.05 |
| $\beta$ | −0.27 ± 0.08 | −0.24 ± 0.11 | −0.29 ± 0.08 | −0.23 ± 0.11 |
| $\gamma$ | 0.34 ± 0.15 | 0.26 ± 0.15 | 0.35 ± 0.15 | 0.26 ± 0.15 |
| $\langle M_{*} L_{B}^{-1} \rangle$ ($M_{\odot} L_{\odot,B}^{-1}$) | 4.8 ± 0.1 | 4.3 ± 0.2 | 2.8 ± 0.1 | 2.8 ± 0.1 |
| $\langle M_{*,0} L_{B,0}^{-1} \rangle$ ($M_{\odot} L_{\odot,B}^{-1}$) | 6.9 ± 0.2 | 6.4 ± 0.2 | 4.0 ± 0.1 | 4.1 ± 0.1 |
| $\delta$ | 0.20 ± 0.02 | 0.23 ± 0.04 | 0.19 ± 0.02 | 0.22 ± 0.04 |
| $\mathrm{d}\log(M_{*} L_{B}^{-1})/\mathrm{d}z$ | −0.71 ± 0.01 | −0.82 ± 0.01 | −0.68 ± 0.01 | −0.82 ± 0.01 |
| $\langle M_{*}(\leq R_{\mathrm{Ein}}) \rangle$ ($10^{10}$ $M_{\odot}$) | 16 ± 1 | 14 ± 1 | 9 ± 1 | 9 ± 1 |
| $\epsilon$ | 0.98 ± 0.05 | 0.98 ± 0.08 | 1.00 ± 0.06 | 0.98 ± 0.08 |
| $\langle f_{*} \rangle$ | 0.71 ± 0.02 | 0.63 ± 0.03 | 0.42 ± 0.01 | 0.41 ± 0.02 |

In Figs. 3 and 4, we show the dependence of both the size ($R_{\mathrm{e}}$) and the surface stellar-mass density [$M_{*}/(2\pi R_{\mathrm{e}}^{2})$] of the galaxies in the sample on their photometric mass. The values of the Pearson linear ($\varrho$) and Kendall rank ($\varsigma$) correlation coefficients (for definitions, see Salkind 2006) are, respectively, higher than 0.84 and 0.60 in the former case and lower than −0.28 and −0.12 in the latter case (see Table 3). It follows that the galaxy size and surface stellar mass density are linearly correlated with the galaxy photometric mass at more than a 95% confidence level. The values of the best-fit power indexes $\alpha$ and $\beta$ of the size-stellar mass

$$R_{\mathrm{e}} \propto M_{*}^{\alpha} \tag{4}$$

and surface stellar mass density-stellar mass

$$\frac{M_{*}}{2\pi R_{\mathrm{e}}^{2}} \propto M_{*}^{\beta} \tag{5}$$

relations are reported in Table 2. We note that similar relations can be created by replacing in the previous two equations the photometric mass by luminosity (e.g., Kormendy 1977).

We remark that our results, derived from a sample of early-type lens galaxies, are consistent, within the errors, with the results derived from a larger sample of SDSS early-type galaxies (see Shen et al. 2003; Cimatti et al. 2008). For instance, Shen et al. (2003) measured a value of 0.56 for $\alpha$ in a sample of more than $10^{5}$ galaxies in the SDSS. In that study, the values of the photometric mass were based on Bruzual & Charlot models and a Kroupa IMF. The scaling laws presented in Eqs. (4) and (5) have been shown to have important implications for the formation and evolution of early-type galaxies in all hierarchical merging models (see Shen et al. 2003; McIntosh et al. 2005; Rettura et al. 2008; van der Wel et al. 2008; Cimatti et al. 2008).

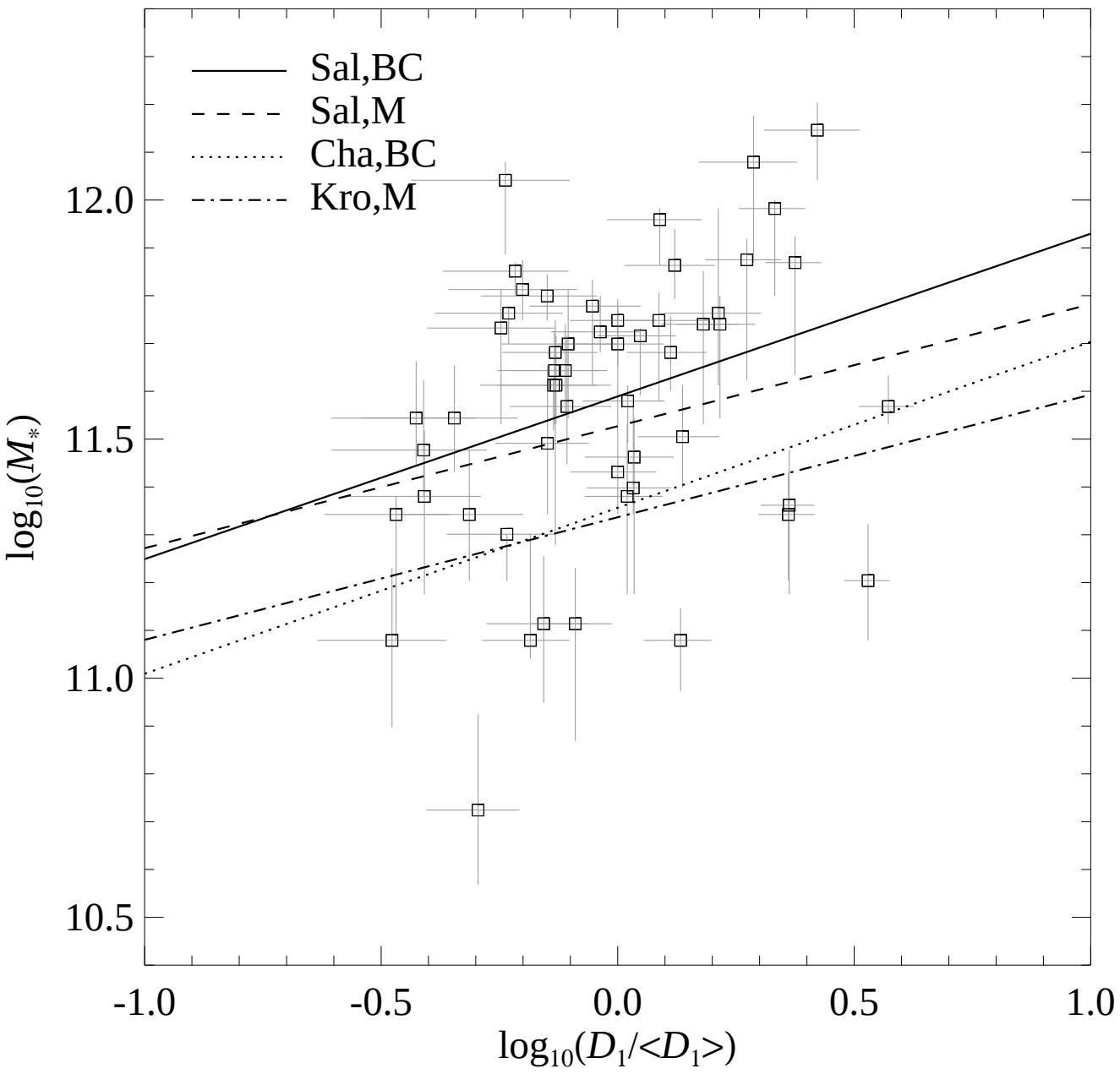

**Fig. 5.** The photometric mass versus the global overdensity, as measured by Treu et al. (2008), of 54 of the 57 early-type grade-A lens galaxies of the SLACS survey. The points with 1 $\sigma$ error bars, obtained by using Bruzual & Charlot models and a Salpeter IMF, and the best-fit correlation lines for all the different composite stellar-population models [Bruzual & Charlot (BC) and Maraston (M)] and IMFs [Salpeter (Sal), Chabrier (Cha), Kroupa (Kro)] are shown.

**Table 3.** The values of the Pearson linear $\varrho$ and Kendall rank $\varsigma$ correlation coefficients.

| | Sal,BC | Sal,M | Cha,BC | Kro,M |
|---|---|---|---|---|
| $\varrho[\log_{10}(R_e), \log_{10}(M_*)]$ | 0.90 (<0.01) | 0.84 (<0.01) | 0.90 (<0.01) | 0.84 (<0.01) |
| $\varsigma[R_e, M_*]$ | 0.67 (<0.01) | 0.60 (<0.01) | 0.67 (<0.01) | 0.60 (<0.01) |
| $\varrho[\log_{10}(M_*/(2\pi R_e^2)), \log_{10}(M_*)]$ | -0.40 (<0.01) | -0.29 (0.03) | -0.43 (<0.01) | -0.28 (0.04) |
| $\varsigma[M_*/(2\pi R_e^2), M_*]$ | -0.20 (0.02) | -0.14 (0.14) | -0.21 (0.02) | -0.12 (0.18) |
| $\varrho[\log_{10}(D_1/\langle D_1\rangle), \log_{10}(M_*)]$ | 0.30 (0.03) | 0.24 (0.09) | 0.31 (0.02) | 0.24 (0.09) |
| $\varsigma[D_1/\langle D_1\rangle, M_*]$ | 0.23 (0.02) | 0.14 (0.15) | 0.23 (0.01) | 0.14 (0.15) |
| $\varrho[\log_{10}(M_{*,0}), \log_{10}(M_{*,0}L_{B,0}^{-1})]$ | 0.74 (<0.01) | 0.65 (<0.01) | 0.72 (<0.01) | 0.64 (<0.01) |
| $\varsigma[M_{*,0}, M_{*,0}L_{B,0}^{-1}]$ | 0.57 (<0.01) | 0.46 (<0.01) | 0.56 (<0.01) | 0.49 (<0.01) |
| $\varrho[z_l, \log_{10}(M_*L_B^{-1}) - \log_{10}(M_{*,0}L_{B,0}^{-1})]$ | −0.87 (<0.01) | −0.87 (<0.01) | −0.90 (<0.01) | −0.88 (<0.01) |
| $\varsigma[z_l, \log_{10}(M_*L_B^{-1}) - \log_{10}(M_{*,0}L_{B,0}^{-1})]$ | −0.69 (<0.01) | −0.75 (<0.01) | −0.74 (<0.01) | −0.76 (<0.01) |
| $\varrho\{\log_{10}[M_*(\leq R_{\rm Ein})], \log_{10}[M_{\rm tot}^{\rm len}(\leq R_{\rm Ein})]\}$ | 0.92 (<0.01) | 0.87 (<0.01) | 0.93 (<0.01) | 0.87 (<0.01) |
| $\varsigma[M_*(\leq R_{\rm Ein}), M_{\rm tot}^{\rm len}(\leq R_{\rm Ein})]$ | 0.74 (<0.01) | 0.68 (<0.01) | 0.74 (<0.01) | 0.69 (<0.01) |

Notes – For the different models and IMFs, the values of the correlation coefficients between the specified variables are given. In parentheses, we show the probabilities that an equal number of measurements of two uncorrelated variables would give values of the coefficients higher than the measured ones.

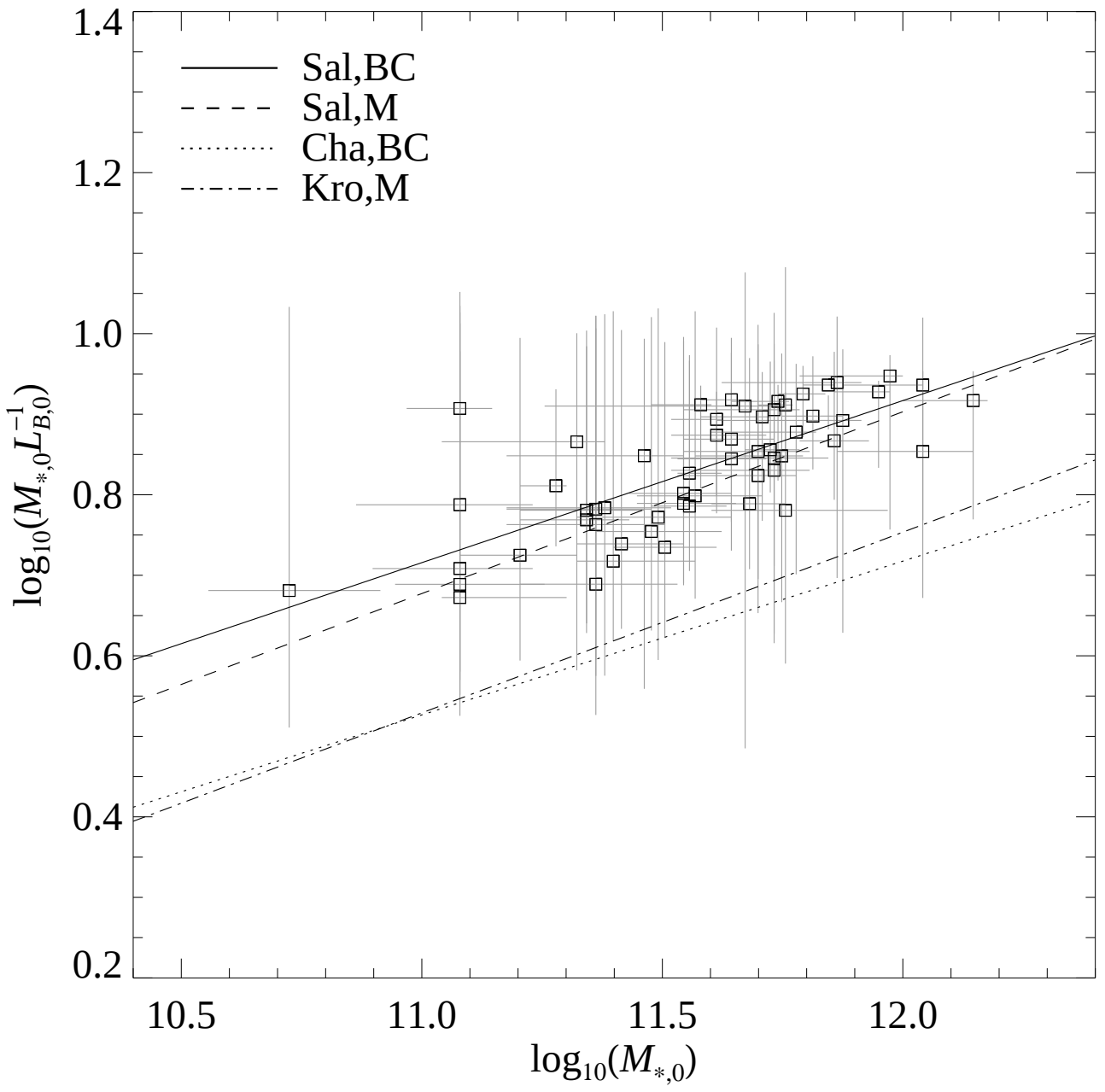

**Fig. 6.** The $B$-band stellar mass-to-light ratio versus the photometric mass at redshift $z = 0$ of the 57 early-type grade-A lens galaxies of the SLACS survey. The points with 1 $\sigma$ error bars, obtained by using Bruzual & Charlot models and a Salpeter IMF, and the best-fit correlation lines for all the different composite stellar-population models [Bruzual & Charlot (BC) and Maraston (M)] and IMFs [Salpeter (Sal), Chabrier (Cha), Kroupa (Kro)] are shown.

We then investigate whether the properties of the environment around the lens galaxies of the sample have any relation to the photometric mass of the galaxies determined by our SED fitting. We consider the environment parameter $D_1$ (defined in Cooper et al. 2005), i.e., the projected number density of galaxies enclosed within a circle centered on each lens of the sample and with radius equal to 1 Mpc. Treu et al. (2008) measured the so-called global overdensity estimator by normalizing $D_1$ to the average density of neighbors $\langle D_1\rangle$ estimated from one hundred random fields of the SDSS. The neighbors are selected by adopting specific magnitude and photometric redshift limits, which depend on the magnitude and redshift of the lens galaxies (for further details, see Treu et al. 2008). The global overdensity was measured for 54 of the 57 galaxies in our sample.

In Fig. 5, we see that the photometric mass $M_*$ is linearly correlated, at a statistically significant level of more than 90% ($\varrho \gtrsim 0.24$ and $\varsigma \gtrsim 0.14$, see Table 3), with the global overdensity $D_1/\langle D_1\rangle$. By fitting the relation

$$M_* \propto \left(\frac{D_1}{\langle D_1\rangle}\right)^{\gamma}, \quad (6)$$

we find the values of the power index $\gamma$ shown in Table 2. Our findings for early-type lens galaxies are consistent with the results from other different studies, according to which early-type galaxies are more luminous and more massive in regions of higher galaxy density (e.g., Dressler et al. 1980).

Next, by assuming a passive evolution from the galaxies' $z_l$ to redshift $z = 0$, we study the dependence of the $B$-band stellar mass-to-light ratio ($M_{*,0}L_{B,0}^{-1}$) on photometric mass ($M_{*,0}$). The data are plotted in Fig. 6.

In particular, we find that the values of the Pearson linear and Kendall rank correlation coefficients are, respectively, higher than 0.64 and 0.46 for $M_{*,0}L_{B,0}^{-1}$ versus $M_{*,0}$ (see Table 3). Thus, the mass-to-light ratio correlates statistically with mass at more than a 99% confidence level. The best-fit values of the exponent $\delta$ of the scaling law

$$\frac{M_{*,0}}{L_{B,0}} \propto M_{*,0}^{\delta} \quad (7)$$

for the different models are given in Table 2.

By assuming that early-type galaxies are homologous stellar systems in virial equilibrium, that their effective (dynamical) mass-to-light ratio $ML^{-1}$ scales as the effective (dynamical) mass $M^{\tilde{\delta}}$ [in a way similar to that expressed in Eq. (7)], and by observing that early-type galaxies are approximately located on the FP, a value of $\tilde{\delta}$ of $0.21 \pm 0.02$ was estimated by Treu et al. (2008) for the Coma galaxies. From a sample of nearly 9000 early-type galaxies of the SDSS between redshift 0.01 and 0.30, it was derived (Bernardi et al. 2003) directly from observed quantities that the dynamical mass-to-light ratio $\sigma_0^2 R_e / L$ scales with dynamical mass as $(\sigma_0^2 R_e)^{0.22 \pm 0.05}$.

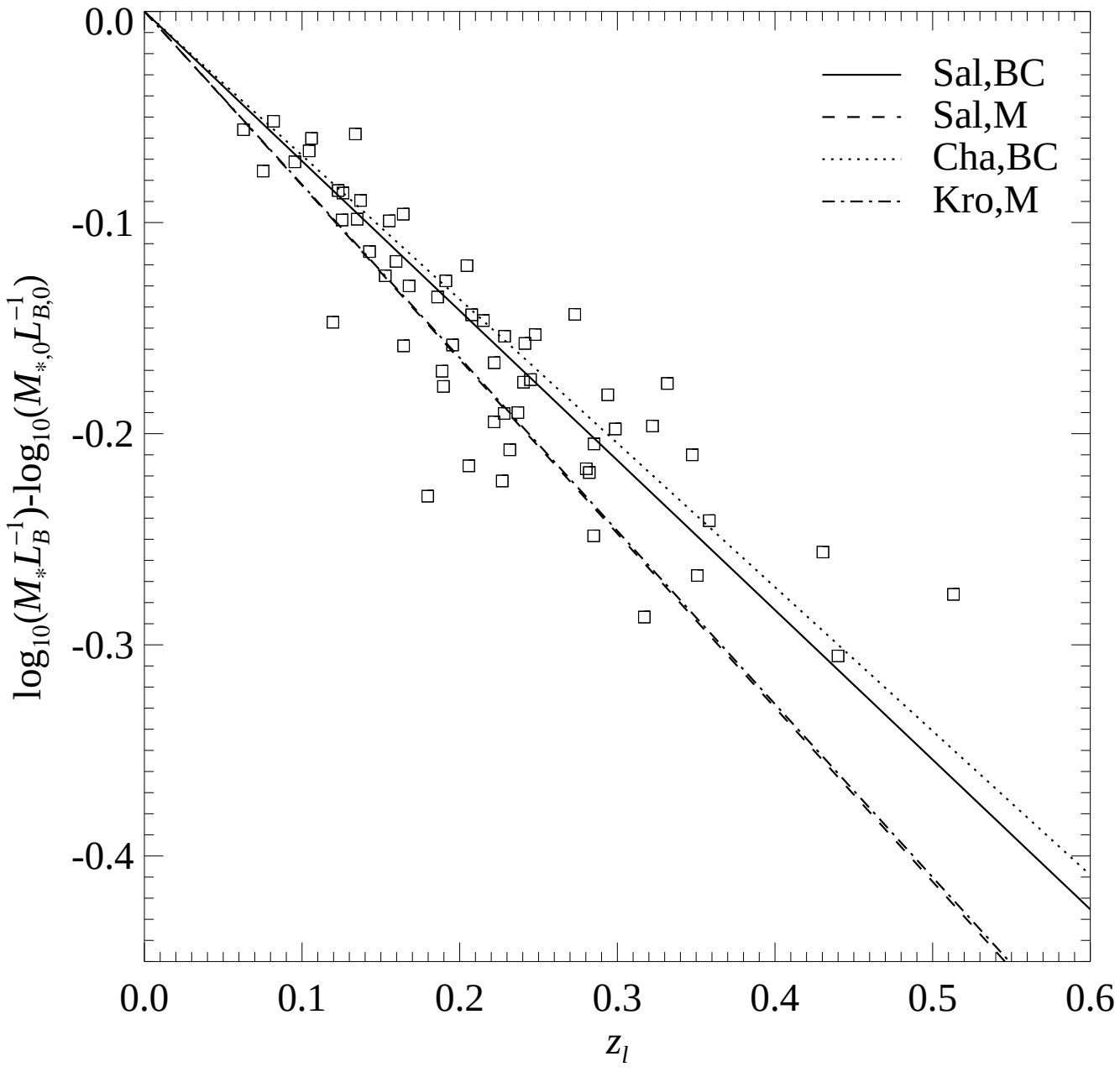

**Fig. 7.** The difference between the logarithms of the stellar mass-to-light ratio at the redshift of the lens galaxies and the stellar mass-to-light ratio at redshift $z = 0$ versus the redshift of the 57 early-type grade-A lens galaxies of the SLACS survey. The points, obtained by using Bruzual & Charlot models and a Salpeter IMF, and the best-fit correlation lines for all the different composite stellar-population models [Bruzual & Charlot (BC) and Maraston (M)] and IMFs [Salpeter (Sal), Chabrier (Cha), Kroupa (Kro)] are shown.

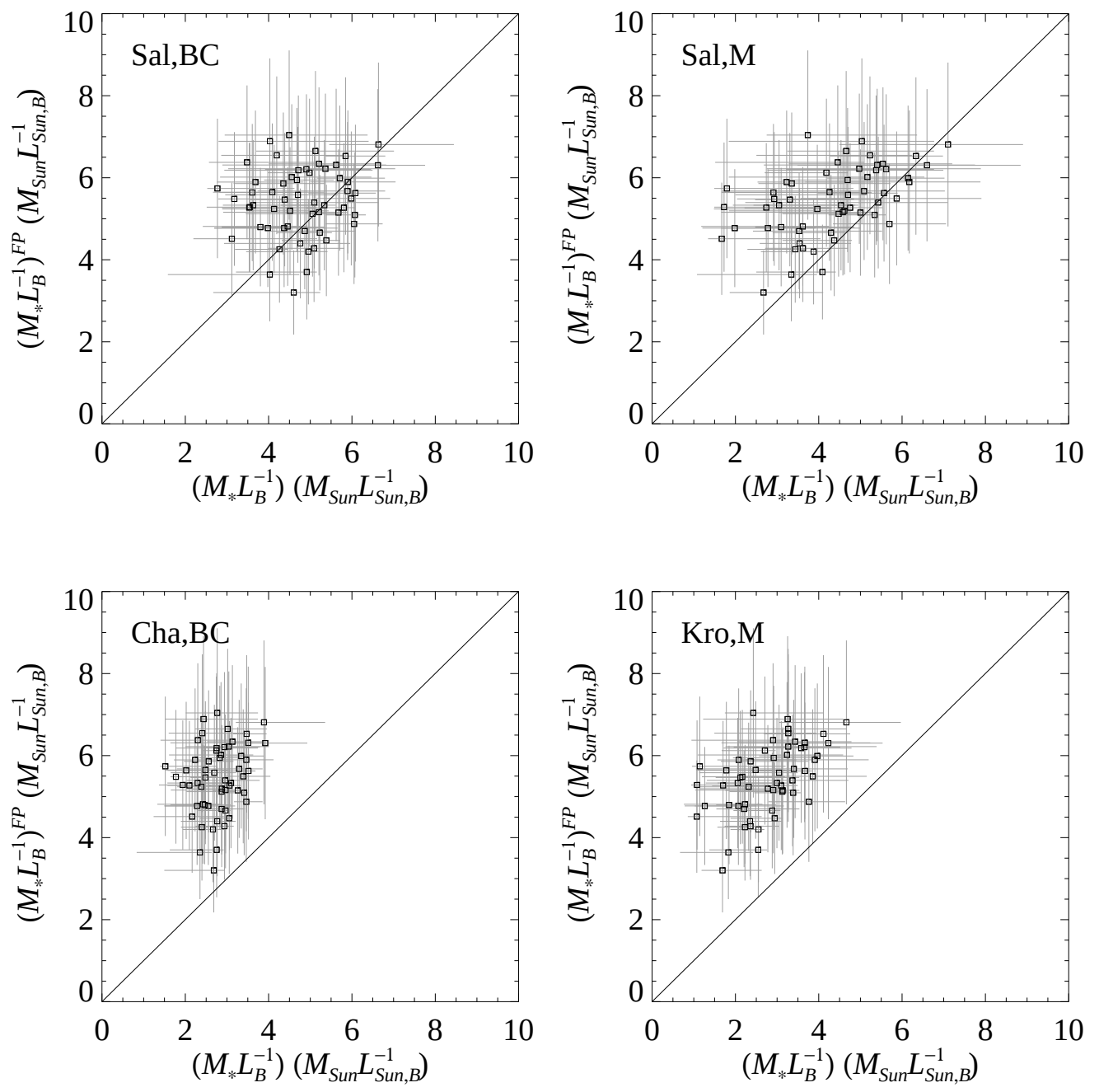

**Fig. 8.** The stellar mass-to-light ratios estimated from the evolution of the FP with redshift versus the stellar mass-to-light ratios obtained by our composite stellar-population models for the 57 early-type grade-A lens galaxies of the SLACS survey (points have 1 $\sigma$ error bars). The mass-to-light ratios from photometry are measured by using different composite stellar-population models [Bruzual & Charlot (BC) and Maraston (M)] and IMFs [Salpeter (Sal), Chabrier (Cha), Kroupa (Kro)]. In each panel, a solid line shows the one-to-one relation.

Our findings about the dependence of the stellar mass-to-light ratio on photometric mass are in good agreement with the cited studies on the corresponding dynamical quantities. It is interesting to notice that the extremely good concordance of the scaling law between mass-to-light ratio and mass, stellar and dynamical, implies that dynamical (or total) mass is linearly proportional to stellar mass. We show further evidence of this result in the next section.

We also investigate the evolution of the stellar mass-to-light ratio with redshift. In Fig. 7, we plot the difference between the logarithms of the stellar mass-to-light ratios measured by the models at the redshift of the lens galaxies and the stellar mass-to-light ratios predicted by assuming passive evolution for our best-fit composite stellar-population models to redshift $z = 0$, $\log(M_* L_B^{-1}) - \log(M_{*,0} L_{B,0}^{-1})$, versus the redshift of the lenses, $z_l$. The correlation of the two variables is statistically significant at a confidence level higher than 99% ($\varrho \lesssim -0.87$ and $\varsigma \lesssim -0.69$, see Table 3), establishing the presence of evolutionary effects in the values of the mass-to-light ratio. By fitting the data given by the different models, we obtain the values of the evolutionary rate, $\mathrm{d}\log(M_* L_B^{-1})/\mathrm{d}z$, that are shown in Table 2.

For appropriate hypotheses (e.g., see Treu et al. 2001, 2005b), the evolution in the intercept of the FP with redshift can be related to the evolution in the average effective mass-to-light ratio. By assuming that the effective mass is then linearly proportional to the stellar mass (i.e., $M \propto M_*$), the evolution in the effective mass-to-light ratio can be connected to the evolution in the stellar mass-to-light ratio. From analyses of this kind, the SLACS survey established for a sample of fifteen galaxies located at redshift lower than $z = 0.33$ that the stellar mass-to-light ratio evolves as $\mathrm{d}\log(M_* L_B^{-1})/\mathrm{d}z = -0.69 \pm 0.08$ (Treu et al. 2006). In the same study, by including five more distant (till redshift $z = 1$) lens galaxies from the LSD survey (Treu & Koopmans 2004), a slightly higher value of $\mathrm{d}\log(M_* L_B^{-1})/\mathrm{d}z = -0.76 \pm 0.03$ was measured. We can conclude that our results based on stellar population models are concordant with those derived from the FP, demonstrating once more the validity of the linear relation between stellar and effective masses.

Finally, in Fig. 8, we compare the absolute values of the stellar mass-to-light ratio at the redshifts of the sample galaxies, as derived from the FP analysis described above and from our best-fit composite stellar-population models. In detail, in the former estimates, the local value of the stellar mass-to-light ratio $(M_{*,0} L_{B,0}^{-1}) = (7.9 \pm 2.3)\, M_\odot L_{\odot,B}^{-1}$ (Gerhard et al.

2001; Treu & Koopmans 2002) and the evolutionary rate value $\mathrm{d}\log(M_* L_B^{-1})/\mathrm{d}z = -0.76 \pm 0.03$ (Treu et al. 2006) are assumed.

We remark that the values of the mass-to-light ratios reproduced by models with a Salpeter IMF are consistent, within the errors, with the values predicted by the FP; in contrast, by adopting Chabrier and Kroupa IMFs, we measure values of the mass-to-light ratios that are systematically lower than those expected from the FP. Moreover, by combining strong and weak gravitational lensing measurements in a sample of 22 early-type lens galaxies of the SLACS survey, Gavazzi et al. (2007) measured an average value of the stellar mass-to-light ratio of $4.2 \pm 0.4\, M_\odot L_{\odot,B}^{-1}$. This value is consistent, within the uncertainties, with our average values derived with a Salpeter IMF, but not with a Chabrier and Kroupa IMF (see Table 2).

## 4. Mass measurements inside the Einstein radii

We describe how the total (luminous+dark) and luminous masses of the lens galaxies in our sample can be measured, within the Einstein ring, by using gravitational lensing and SED fitting methods.

### 4.1. Total masses from gravitational lensing

The distribution of the total mass of the galaxies in the sample was successfully modeled by Bolton et al. (2008a) in terms of a one-component singular isothermal ellipsoid (SIE). This model can be parametrized by the values of a length scale (related to the projected distance of the multiple images of a lensing system and called the Einstein radius in the axisymmetric case), ellipticity, and position angle. By choosing the normalization adopted by Korman et al. (1994), the value of the projected mass enclosed within the elliptical critical curves of an SIE is independent of the value of the ellipticity of the model. Thus, the value of the projected mass enclosed within the tangential critical curve of an SIE model can be expressed as in an "effective" axisymmetric model as

$$M_{\rm tot}^{\rm len}(\le R_{\rm Ein}) = \Sigma_{\rm cr} \pi R_{\rm Ein}^2 \,, \tag{8}$$

where $R_{\rm Ein}$ is the "effective" Einstein radius of the lensing system and $\Sigma_{\rm cr}$ is a geometrical factor, called critical surface-mass density. This factor depends only on the angular diameter distances between observer, lens, and source ($D_{\rm ol}$, $D_{\rm ls}$, and $D_{\rm os}$), and is defined as

$$\Sigma_{\rm cr} = \frac{c^2}{4\pi G} \frac{D_{\rm os}}{D_{\rm ol} D_{\rm ls}} \,, \tag{9}$$

where $c$ is the light speed and $G$ is the gravitational constant.

We show in Table 4 the values of the total projected mass and their errors for the galaxies in our sample. These values are estimated by applying Eq. (8) to the data of Table 1 and assuming a conservative 5% error in the values of the "effective" Einstein angles measured by Bolton et al. (2008a). For the sample galaxies, the average value of the total mass projected within the average Einstein radius of approximately 4 kpc is $\langle M_{\rm tot}^{\rm len}(\le R_{\rm Ein})\rangle = (23 \pm 2) \times 10^{10} M_\odot$.

### 4.2. Luminous masses from photometry

The photometric mass estimates of the galaxies in our sample are performed by fitting their SEDs, as explained in the previous section. The mass in the form of stars enclosed within the disk defined by the Einstein radius of each lensing system [$M_*^{\rm Sal,BC}(\le R_{\rm Ein})$, $M_*^{\rm Sal,M}(\le R_{\rm Ein})$, $M_*^{\rm Cha,BC}(\le R_{\rm Ein})$, and $M_*^{\rm Kro,M}(\le R_{\rm Ein})$] is inferred then for each galaxy by multiplying the best-fit value of the photometric mass by an aperture factor ($f_{\rm ap}$)

$$M_*(\le R_{\rm Ein}) = f_{\rm ap}\, M_* \,. \tag{10}$$

This last factor is defined as the light measured within the Einstein ring divided by the total light of the galaxy parametrized by a de Vaucouleurs luminosity profile. In detail, the analytical expression is given by

$$f_{\rm ap} = \frac{\int_0^{R_{\rm Ein}} I(R)\, R\, \mathrm{d}R}{\int_0^{\infty} I(R)\, R\, \mathrm{d}R} \,, \tag{11}$$

where $I(R)$ is given by Eq. (1). The implicit assumption made in this calculation is that the stellar mass is traced by the light distribution.

In the sample, the mean value of the aperture factor is $\langle f_{\rm ap}\rangle = 0.36 \pm 0.01$. Thus, on average, more that one-third of the light (luminous mass) is contained within the disk defined by the Einstein radius of the lenses. In Table 4, we report the estimates of the mass in the form of stars inside the Einstein radii of the lens galaxies for the different composite stellar-population models and IMFs. These mass values range between $1 \times 10^{10}\, M_\odot$ and $40 \times 10^{10}\, M_\odot$ and their average values are shown in Table 2. We note that the previously mentioned dependence of the photometric mass measurements on the adopted IMF is clearly reflected here.

### 4.3. Luminous and dark matter

To study the relationship between luminous and dark matter, we analyze the lensing (total, i.e., luminous and dark) and photometric (luminous) masses of the 57 grade-A lens early-type galaxies of the SLACS survey (see Table 4) in a statistical way, by assuming that the structure of all the lenses is homologous.

In Fig. 9, we plot the total versus luminous-mass estimates for the different models and IMFs under investigation. We find that the two quantities are statistically correlated at more than a 99% confidence level, as demonstrated by the values of the Pearson linear and Kendall rank correlation coefficients (see Table 3), which are higher than 0.87 and 0.68, respectively, in all four cases. From the plot, we note that the best-fit correlation line of all the models is well-approximated by a line parallel to the one-to-one relation line. In particular, we show in Table 2 that the best-fit values of the exponent $\epsilon$ of the relation

$$M_{\rm tot}^{\rm len}(\le R_{\rm Ein}) \propto [M_*(\le R_{\rm Ein})]^{\epsilon} \tag{12}$$

are consistent with unity. Two qualitatively different scenarios are evident by looking first at the top panels and then at the bottom ones. The intercepts of the best-fit correlation lines in the

**Table 4.** The photometric and lensing mass measurements of the 57 early-type grade-A lens galaxies of the SLACS survey.

| SDSS Name | $M_*^{\rm Sal,BC}$ ($10^{10}$ M$_\odot$) | $M_*^{\rm Sal,M}$ ($10^{10}$ M$_\odot$) | $M_*^{\rm Cha,BC}$ ($10^{10}$ M$_\odot$) | $M_*^{\rm Kro,M}$ ($10^{10}$ M$_\odot$) | $f_{ap}$ | $M_*^{\rm Sal,BC}(\leq R_{\rm Ein})$ ($10^{10}$ M$_\odot$) | $M_*^{\rm Sal,M}(\leq R_{\rm Ein})$ ($10^{10}$ M$_\odot$) | $M_*^{\rm Cha,BC}(\leq R_{\rm Ein})$ ($10^{10}$ M$_\odot$) | $M_*^{\rm Kro,M}(\leq R_{\rm Ein})$ ($10^{10}$ M$_\odot$) | $M_{\rm tot}^{\rm len}(\leq R_{\rm Ein})$ ($10^{10}$ M$_\odot$) |
|---|---|---|---|---|---|---|---|---|---|---|
| J0008−0004 | $48^{+8}_{-29}$ | $40^{+7}_{-27}$ | $28^{+3}_{-18}$ | $22^{+7}_{-14}$ | 0.40 | $19^{+3}_{-11}$ | $16^{+3}_{-11}$ | $11^{+1}_{-7}$ | $9^{+3}_{-6}$ | $35 \pm 4$ |
| J0029−0055 | $31^{+14}_{-9}$ | $15^{+22}_{-2}$ | $17^{+8}_{-5}$ | $9^{+15}_{-1}$ | 0.29 | $9^{+4}_{-3}$ | $4^{+6}_{-1}$ | $5^{+2}_{-1}$ | $3^{+4}_{-1}$ | $12 \pm 1$ |
| J0037−0942 | $54^{+11}_{-20}$ | $54^{+5}_{-31}$ | $31^{+6}_{-12}$ | $35^{+5}_{-22}$ | 0.40 | $22^{+4}_{-8}$ | $22^{+2}_{-13}$ | $13^{+2}_{-5}$ | $14^{+2}_{-9}$ | $29 \pm 3$ |
| J0044+0113 | $23^{+11}_{-6}$ | $29^{+4}_{-19}$ | $15^{+5}_{-6}$ | $19^{+3}_{-13}$ | 0.22 | $5^{+2}_{-1}$ | $6^{+1}_{-4}$ | $3^{+1}_{-1}$ | $4^{+1}_{-3}$ | $9 \pm 1$ |
| J0109+1500 | $44^{+11}_{-8}$ | $32^{+18}_{-10}$ | $26^{+5}_{-6}$ | $20^{+12}_{-6}$ | 0.32 | $14^{+4}_{-3}$ | $10^{+6}_{-3}$ | $8^{+2}_{-2}$ | $6^{+4}_{-2}$ | $13 \pm 1$ |
| J0157−0056 | $74^{+10}_{-31}$ | $43^{+23}_{-13}$ | $43^{+4}_{-19}$ | $27^{+15}_{-8}$ | 0.42 | $31^{+4}_{-13}$ | $18^{+10}_{-5}$ | $18^{+2}_{-8}$ | $11^{+6}_{-3}$ | $26 \pm 3$ |
| J0216−0813 | $120^{+30}_{-46}$ | $89^{+31}_{-24}$ | $70^{+11}_{-22}$ | $59^{+19}_{-18}$ | 0.29 | $35^{+9}_{-13}$ | $26^{+9}_{-7}$ | $20^{+3}_{-6}$ | $17^{+5}_{-5}$ | $49 \pm 5$ |
| J0252+0039 | $22^{+2}_{-10}$ | $18^{+3}_{-12}$ | $12^{+2}_{-5}$ | $11^{+2}_{-7}$ | 0.42 | $9^{+1}_{-4}$ | $8^{+1}_{-5}$ | $5^{+1}_{-2}$ | $5^{+1}_{-3}$ | $18 \pm 2$ |
| J0330−0020 | $52^{+2}_{-13}$ | $42^{+3}_{-14}$ | $29^{+2}_{-6}$ | $27^{+1}_{-8}$ | 0.48 | $25^{+1}_{-6}$ | $20^{+1}_{-7}$ | $14^{+1}_{-3}$ | $13^{+1}_{-4}$ | $25 \pm 3$ |
| J0405−0455 | $5^{+3}_{-2}$ | $7^{+2}_{-4}$ | $3^{+2}_{-1}$ | $4^{+2}_{-3}$ | 0.36 | $2^{+1}_{-1}$ | $2^{+1}_{-1}$ | $1^{+1}_{-1}$ | $2^{+1}_{-1}$ | $3 \pm 1$ |
| J0728+3835 | $25^{+12}_{-3}$ | $23^{+15}_{-7}$ | $14^{+8}_{-1}$ | $17^{+8}_{-8}$ | 0.41 | $10^{+5}_{-1}$ | $9^{+6}_{-3}$ | $6^{+3}_{-1}$ | $7^{+3}_{-3}$ | $20 \pm 2$ |
| J0737+3216 | $91^{+5}_{-18}$ | $74^{+7}_{-21}$ | $52^{+1}_{-12}$ | $50^{+3}_{-18}$ | 0.25 | $22^{+1}_{-4}$ | $18^{+2}_{-5}$ | $13^{+1}_{-3}$ | $12^{+1}_{-4}$ | $29 \pm 3$ |
| J0822+2652 | $58^{+2}_{-8}$ | $51^{+3}_{-20}$ | $33^{+1}_{-6}$ | $32^{+4}_{-14}$ | 0.38 | $22^{+1}_{-3}$ | $20^{+1}_{-8}$ | $13^{+1}_{-2}$ | $12^{+2}_{-5}$ | $24 \pm 2$ |
| J0903+4116 | $96^{+4}_{-33}$ | $80^{+6}_{-31}$ | $54^{+3}_{-22}$ | $50^{+4}_{-23}$ | 0.41 | $40^{+2}_{-14}$ | $33^{+2}_{-13}$ | $22^{+1}_{-9}$ | $21^{+2}_{-10}$ | $45 \pm 5$ |
| J0912+0029 | $73^{+14}_{-11}$ | $77^{+21}_{-31}$ | $43^{+8}_{-8}$ | $49^{+16}_{-21}$ | 0.28 | $21^{+4}_{-3}$ | $22^{+6}_{-9}$ | $12^{+2}_{-2}$ | $14^{+5}_{-6}$ | $40 \pm 4$ |
| J0935−0003 | $140^{+20}_{-30}$ | $100^{+30}_{-25}$ | $81^{+6}_{-19}$ | $65^{+19}_{-17}$ | 0.15 | $21^{+3}_{-5}$ | $15^{+5}_{-4}$ | $12^{+1}_{-3}$ | $10^{+3}_{-3}$ | $41 \pm 4$ |
| J0936+0913 | $32^{+9}_{-7}$ | $26^{+21}_{-11}$ | $18^{+5}_{-4}$ | $16^{+14}_{-6}$ | 0.33 | $11^{+3}_{-2}$ | $9^{+7}_{-4}$ | $6^{+2}_{-1}$ | $5^{+5}_{-2}$ | $15 \pm 1$ |
| J0946+1006 | $41^{+2}_{-8}$ | $35^{+3}_{-20}$ | $24^{+1}_{-6}$ | $23^{+2}_{-13}$ | 0.36 | $15^{+1}_{-3}$ | $13^{+1}_{-7}$ | $9^{+1}_{-2}$ | $8^{+1}_{-5}$ | $29 \pm 3$ |
| J0956+5100 | $75^{+8}_{-33}$ | $66^{+11}_{-42}$ | $43^{+4}_{-19}$ | $42^{+6}_{-27}$ | 0.37 | $28^{+3}_{-12}$ | $24^{+4}_{-15}$ | $16^{+1}_{-7}$ | $15^{+2}_{-10}$ | $37 \pm 4$ |
| J0959+4416 | $44^{+9}_{-10}$ | $45^{+1}_{-27}$ | $28^{+1}_{-7}$ | $27^{+3}_{-16}$ | 0.31 | $14^{+3}_{-3}$ | $14^{+1}_{-8}$ | $9^{+1}_{-2}$ | $8^{+1}_{-5}$ | $17 \pm 2$ |
| J0959+0410 | $12^{+2}_{-3}$ | $12^{+4}_{-6}$ | $7^{+2}_{-2}$ | $8^{+2}_{-4}$ | 0.41 | $5^{+1}_{-1}$ | $5^{+2}_{-2}$ | $3^{+1}_{-1}$ | $3^{+1}_{-2}$ | $8 \pm 1$ |
| J1016+3859 | $22^{+5}_{-6}$ | $17^{+12}_{-4}$ | $13^{+3}_{-3}$ | $12^{+7}_{-4}$ | 0.42 | $9^{+2}_{-3}$ | $7^{+5}_{-2}$ | $5^{+1}_{-1}$ | $5^{+3}_{-2}$ | $15 \pm 1$ |
| J1020+1122 | $37^{+15}_{-9}$ | $30^{+17}_{-6}$ | $24^{+6}_{-8}$ | $18^{+11}_{-3}$ | 0.42 | $16^{+6}_{-4}$ | $13^{+7}_{-3}$ | $10^{+3}_{-3}$ | $8^{+5}_{-1}$ | $34 \pm 3$ |
| J1023+4230 | $38^{+3}_{-7}$ | $35^{+2}_{-10}$ | $22^{+1}_{-4}$ | $23^{+1}_{-8}$ | 0.44 | $17^{+1}_{-3}$ | $15^{+1}_{-4}$ | $10^{+1}_{-2}$ | $10^{+1}_{-4}$ | $23 \pm 2$ |
| J1029+0420 | $13^{+5}_{-4}$ | $16^{+3}_{-8}$ | $8^{+4}_{-3}$ | $10^{+2}_{-5}$ | 0.38 | $5^{+2}_{-2}$ | $6^{+1}_{-3}$ | $3^{+1}_{-1}$ | $4^{+1}_{-2}$ | $6 \pm 1$ |
| J1100+5329 | $58^{+38}_{-17}$ | $31^{+44}_{-6}$ | $40^{+14}_{-17}$ | $20^{+29}_{-4}$ | 0.40 | $23^{+15}_{-7}$ | $12^{+17}_{-2}$ | $16^{+6}_{-7}$ | $8^{+11}_{-2}$ | $47 \pm 5$ |
| J1106+5228 | $22^{+8}_{-6}$ | $20^{+9}_{-6}$ | $13^{+4}_{-3}$ | $14^{+6}_{-5}$ | 0.42 | $9^{+3}_{-2}$ | $8^{+4}_{-2}$ | $5^{+2}_{-1}$ | $6^{+2}_{-2}$ | $9 \pm 1$ |
| J1112+0826 | $63^{+7}_{-7}$ | $59^{+14}_{-14}$ | $36^{+4}_{-3}$ | $39^{+9}_{-10}$ | 0.50 | $31^{+3}_{-3}$ | $29^{+7}_{-7}$ | $18^{+2}_{-1}$ | $19^{+4}_{-5}$ | $45 \pm 4$ |
| J1134+6027 | $24^{+10}_{-9}$ | $27^{+7}_{-17}$ | $15^{+4}_{-7}$ | $17^{+7}_{-11}$ | 0.34 | $8^{+3}_{-3}$ | $9^{+2}_{-6}$ | $5^{+1}_{-2}$ | $6^{+2}_{-4}$ | $13 \pm 1$ |
| J1142+1001 | $30^{+12}_{-6}$ | $25^{+15}_{-5}$ | $19^{+6}_{-6}$ | $17^{+9}_{-4}$ | 0.33 | $10^{+4}_{-2}$ | $8^{+5}_{-2}$ | $6^{+2}_{-2}$ | $6^{+3}_{-1}$ | $17 \pm 2$ |
| J1143−0144 | $37^{+6}_{-3}$ | $40^{+4}_{-13}$ | $22^{+3}_{-2}$ | $26^{+4}_{-9}$ | 0.24 | $9^{+1}_{-1}$ | $10^{+1}_{-3}$ | $5^{+1}_{-1}$ | $6^{+1}_{-2}$ | $19 \pm 2$ |
| J1153+4612 | $12^{+8}_{-1}$ | $8^{+13}_{-1}$ | $7^{+5}_{-1}$ | $5^{+9}_{-1}$ | 0.47 | $6^{+4}_{-1}$ | $4^{+6}_{-1}$ | $3^{+3}_{-1}$ | $2^{+4}_{-1}$ | $11 \pm 1$ |
| J1204+0358 | $16^{+5}_{-4}$ | $14^{+9}_{-3}$ | $10^{+2}_{-2}$ | $9^{+6}_{-2}$ | 0.47 | $8^{+2}_{-2}$ | $7^{+4}_{-1}$ | $5^{+1}_{-1}$ | $4^{+3}_{-1}$ | $17 \pm 2$ |
| J1205+4910 | $50^{+15}_{-15}$ | $53^{+16}_{-31}$ | $29^{+8}_{-9}$ | $33^{+13}_{-20}$ | 0.31 | $15^{+5}_{-5}$ | $16^{+5}_{-10}$ | $9^{+2}_{-3}$ | $10^{+4}_{-6}$ | $25 \pm 3$ |
| J1213+6708 | $35^{+10}_{-8}$ | $37^{+11}_{-15}$ | $21^{+5}_{-4}$ | $23^{+9}_{-9}$ | 0.29 | $10^{+3}_{-2}$ | $11^{+3}_{-4}$ | $6^{+1}_{-1}$ | $7^{+3}_{-3}$ | $14 \pm 1$ |
| J1218+0830 | $35^{+11}_{-7}$ | $40^{+8}_{-17}$ | $21^{+6}_{-4}$ | $26^{+6}_{-12}$ | 0.30 | $10^{+3}_{-2}$ | $12^{+2}_{-5}$ | $6^{+2}_{-1}$ | $8^{+2}_{-4}$ | $16 \pm 2$ |
| J1250+0523 | $50^{+12}_{-15}$ | $48^{+4}_{-30}$ | $29^{+5}_{-7}$ | $28^{+7}_{-16}$ | 0.38 | $19^{+5}_{-6}$ | $18^{+2}_{-11}$ | $11^{+2}_{-3}$ | $11^{+3}_{-6}$ | $18 \pm 2$ |
| J1402+6321 | $65^{+10}_{-9}$ | $64^{+22}_{-25}$ | $37^{+7}_{-5}$ | $42^{+14}_{-18}$ | 0.32 | $21^{+3}_{-3}$ | $21^{+7}_{-8}$ | $12^{+2}_{-2}$ | $13^{+4}_{-6}$ | $29 \pm 3$ |
| J1403+0006 | $23^{+7}_{-8}$ | $24^{+4}_{-15}$ | $14^{+3}_{-6}$ | $14^{+4}_{-8}$ | 0.35 | $8^{+2}_{-3}$ | $8^{+1}_{-5}$ | $5^{+1}_{-2}$ | $5^{+1}_{-3}$ | $10 \pm 1$ |
| J1416+5136 | $56^{+3}_{-11}$ | $46^{+4}_{-15}$ | $32^{+1}_{-7}$ | $31^{+2}_{-10}$ | 0.49 | $27^{+1}_{-5}$ | $22^{+2}_{-7}$ | $16^{+1}_{-3}$ | $15^{+1}_{-5}$ | $37 \pm 4$ |
| J1420+6019 | $12^{+5}_{-4}$ | $10^{+7}_{-3}$ | $7^{+3}_{-2}$ | $7^{+5}_{-2}$ | 0.32 | $4^{+2}_{-1}$ | $3^{+2}_{-1}$ | $2^{+1}_{-1}$ | $2^{+1}_{-1}$ | $4 \pm 1$ |
| J1430+4105 | $56^{+8}_{-18}$ | $39^{+16}_{-22}$ | $32^{+4}_{-11}$ | $29^{+6}_{-18}$ | 0.36 | $20^{+3}_{-7}$ | $14^{+6}_{-8}$ | $12^{+1}_{-4}$ | $11^{+2}_{-7}$ | $54 \pm 5$ |
| J1436−0000 | $55^{+16}_{-21}$ | $25^{+38}_{-4}$ | $32^{+8}_{-13}$ | $16^{+24}_{-3}$ | 0.32 | $18^{+5}_{-7}$ | $8^{+12}_{-1}$ | $10^{+3}_{-4}$ | $5^{+8}_{-1}$ | $23 \pm 2$ |
| J1443+0304 | $13^{+4}_{-6}$ | $12^{+7}_{-5}$ | $7^{+2}_{-3}$ | $8^{+5}_{-4}$ | 0.46 | $6^{+2}_{-3}$ | $6^{+3}_{-2}$ | $3^{+1}_{-1}$ | $4^{+2}_{-2}$ | $6 \pm 1$ |
| J1451−0239 | $29^{+6}_{-14}$ | $28^{+12}_{-14}$ | $18^{+2}_{-10}$ | $19^{+8}_{-10}$ | 0.28 | $8^{+2}_{-4}$ | $8^{+3}_{-4}$ | $5^{+1}_{-3}$ | $5^{+2}_{-3}$ | $8 \pm 1$ |
| J1525+3327 | $110^{+10}_{-33}$ | $86^{+8}_{-28}$ | $59^{+6}_{-16}$ | $57^{+3}_{-15}$ | 0.30 | $33^{+3}_{-10}$ | $26^{+2}_{-8}$ | $18^{+2}_{-5}$ | $17^{+1}_{-4}$ | $48 \pm 5$ |
| J1531−0105 | $48^{+9}_{-8}$ | $48^{+15}_{-16}$ | $29^{+5}_{-5}$ | $30^{+11}_{-10}$ | 0.40 | $19^{+4}_{-3}$ | $19^{+6}_{-6}$ | $12^{+2}_{-2}$ | $12^{+4}_{-4}$ | $27 \pm 3$ |
| J1538+5817 | $20^{+1}_{-4}$ | $17^{+2}_{-6}$ | $11^{+1}_{-1}$ | $11^{+1}_{-4}$ | 0.38 | $8^{+1}_{-2}$ | $6^{+1}_{-2}$ | $4^{+1}_{-1}$ | $4^{+1}_{-2}$ | $9 \pm 1$ |
| J1621+3931 | $60^{+8}_{-19}$ | $53^{+6}_{-16}$ | $34^{+4}_{-10}$ | $37^{+1}_{-15}$ | 0.37 | $22^{+3}_{-7}$ | $19^{+2}_{-6}$ | $12^{+1}_{-4}$ | $14^{+1}_{-5}$ | $29 \pm 3$ |
| J1627−0053 | $37^{+6}_{-7}$ | $28^{+18}_{-7}$ | $21^{+4}_{-3}$ | $18^{+12}_{-5}$ | 0.37 | $14^{+2}_{-3}$ | $10^{+7}_{-3}$ | $8^{+1}_{-1}$ | $7^{+4}_{-2}$ | $23 \pm 2$ |
| J1630+4520 | $71^{+3}_{-8}$ | $63^{+3}_{-13}$ | $40^{+1}_{-4}$ | $40^{+2}_{-11}$ | 0.47 | $34^{+1}_{-4}$ | $30^{+1}_{-6}$ | $19^{+1}_{-2}$ | $19^{+1}_{-5}$ | $49 \pm 5$ |
| J1636+4707 | $27^{+8}_{-5}$ | $21^{+12}_{-4}$ | $16^{+4}_{-4}$ | $13^{+10}_{-2}$ | 0.38 | $10^{+3}_{-2}$ | $8^{+5}_{-2}$ | $6^{+2}_{-2}$ | $5^{+4}_{-1}$ | $18 \pm 2$ |
| J2238−0754 | $24^{+9}_{-9}$ | $27^{+8}_{-13}$ | $14^{+5}_{-5}$ | $18^{+6}_{-9}$ | 0.34 | $8^{+3}_{-3}$ | $9^{+3}_{-4}$ | $5^{+2}_{-2}$ | $6^{+2}_{-3}$ | $13 \pm 1$ |
| J2300+0022 | $45^{+2}_{-7}$ | $37^{+6}_{-14}$ | $24^{+2}_{-4}$ | $24^{+3}_{-10}$ | 0.40 | $18^{+1}_{-3}$ | $15^{+2}_{-6}$ | $10^{+1}_{-2}$ | $10^{+1}_{-4}$ | $30 \pm 3$ |
| J2303+1422 | $53^{+10}_{-5}$ | $57^{+8}_{-18}$ | $31^{+5}_{-3}$ | $37^{+8}_{-13}$ | 0.32 | $17^{+3}_{-2}$ | $18^{+3}_{-6}$ | $10^{+2}_{-1}$ | $12^{+3}_{-4}$ | $27 \pm 3$ |
| J2321−0939 | $41^{+11}_{-7}$ | $44^{+11}_{-16}$ | $24^{+9}_{-4}$ | $29^{+8}_{-10}$ | 0.26 | $11^{+3}_{-2}$ | $12^{+3}_{-4}$ | $6^{+2}_{-1}$ | $8^{+2}_{-3}$ | $12 \pm 1$ |
| J2341+0000 | $55^{+8}_{-20}$ | $48^{+18}_{-32}$ | $31^{+5}_{-14}$ | $32^{+12}_{-22}$ | 0.30 | $16^{+2}_{-6}$ | $14^{+5}_{-10}$ | $9^{+1}_{-4}$ | $10^{+4}_{-7}$ | $22 \pm 2$ |

top panels are indeed similar, so that the two best-fit correlation lines of the first models are almost identical. The same can be concluded by considering the two bottom panels. The mean values, and their errors, of the ratios of the total to luminous masses of the four models are $1.5 \pm 0.1$, $1.8 \pm 0.1$, $2.6 \pm 0.1$, and $2.7 \pm 0.1$, respectively. According to all models, these results suggest that the total mass is linearly proportional to the mass in the form of stars for the lenses in the sample, that the photometric mass estimates of the first (last) two models are consistent (as discussed above), and that, on average, the dark matter component, which must be added to the luminous component to provide the total mass of the lenses of the sample, is considerably higher for the last two models than for the first ones.

We then evaluate the fraction of projected mass in the form of stars enclosed within the disk defined by the Einstein radius of each lensing system as the ratio of the luminous to the total projected mass of each model ($f_*^{\rm Sal,BC}$, $f_*^{\rm Sal,M}$, $f_*^{\rm Cha,BC}$, and $f_*^{\rm Kro,M}$).

We obtain the values presented in Tables 2 and 5 and plotted in Fig. 10 versus an adimensional radius given by the ratio of the Einstein radius to the effective radius of each lens. This adi-mensional radius describes how close the investigated region is to the center of a lens more conveniently than the Einstein radius.

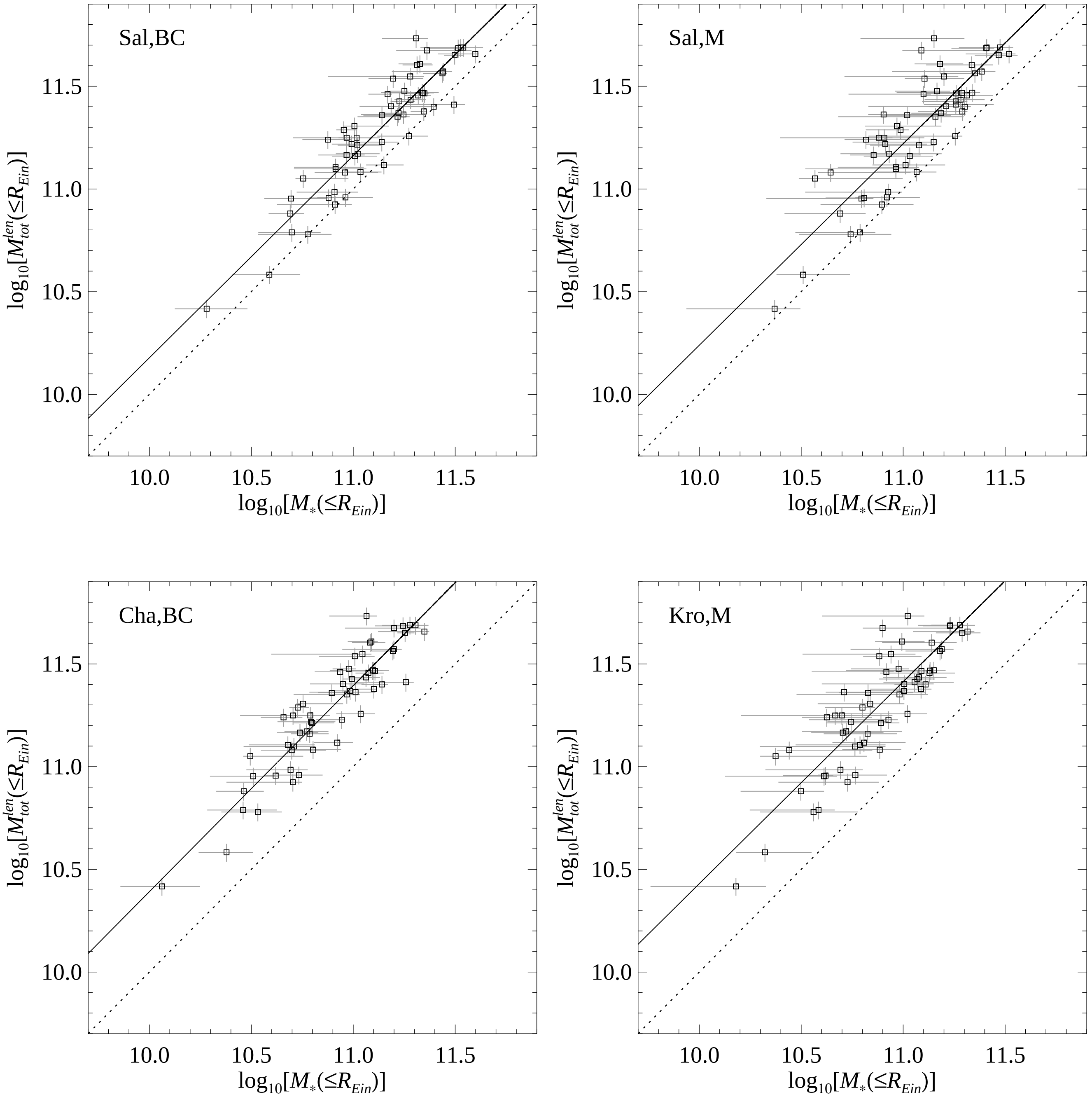

**Fig. 9.** The total mass derived from strong gravitational lensing versus the mass contained in the form of stars obtained by different composite stellar-population models [Bruzual & Charlot (BC) and Maraston (M)] and IMFs [Salpeter (Sal), Chabrier (Cha), Kroupa (Kro)] for the 57 early-type grade-A lens galaxies of the SLACS survey. The mass estimates (with 1 $\sigma$ error bars) are considered within the disks defined by the Einstein radii of the lensing systems. In each panel, the best-fit correlation and the one-to-one relation lines are shown by solid and dotted lines respectively.

Surprisingly, in Fig. 10, we do not find any statistically significant evidence of a decrease in the fraction of mass in the form of stars between 0.2 and 1.0 $R_e$ in any of the models. Although it is less straightforward than here, the same result can also be found in the analysis of the first 15 galaxies of the SLACS survey performed by Koopmans et al. (2006) (see Fig. 2). In that joint gravitational lensing and stellar-dynamical study, an average value of $0.25\pm0.06$ was measured for the dark matter fraction projected inside the average Einstein radius of $4.2 \pm 0.4$ kpc. A slightly higher value ($0.37 \pm 0.04$) of the same quantity is estimated by Gavazzi et al. (2007) from weak- and

**Table 5.** The fractions of mass in the form of stars and the mass-to-light ratios of the 57 early-type grade-A lens galaxies of the SLACS survey.

| SDSS Name | $f_*^{\rm Sal,BC}$ | $f_*^{\rm Sal,M}$ | $f_*^{\rm Cha,BC}$ | $f_*^{\rm Kro,M}$ | $(M_* L_B^{-1})^{\rm Sal,BC}$ $(M_\odot L_{\odot,B}^{-1})$ | $(M_* L_B^{-1})^{\rm Sal,M}$ $(M_\odot L_{\odot,B}^{-1})$ | $(M_* L_B^{-1})^{\rm Cha,BC}$ $(M_\odot L_{\odot,B}^{-1})$ | $(M_* L_B^{-1})^{\rm Kro,M}$ $(M_\odot L_{\odot,B}^{-1})$ | $(M_* L_B^{-1})^{\rm FP}$ $(M_\odot L_{\odot,B}^{-1})$ |
|---|---|---|---|---|---|---|---|---|---|
| J0008−0004 | $0.54^{+0.10}_{-0.33}$ | $0.45^{+0.09}_{-0.31}$ | $0.31^{+0.05}_{-0.20}$ | $0.25^{+0.08}_{-0.16}$ | $4.0^{+0.7}_{-2.4}$ | $3.3^{+0.6}_{-2.3}$ | $2.4^{+0.3}_{-1.5}$ | $1.8^{+0.6}_{-1.2}$ | $3.6 \pm 1.1$ |
| J0029−0055 | $0.76^{+0.35}_{-0.23}$ | $0.37^{+0.54}_{-0.06}$ | $0.42^{+0.20}_{-0.13}$ | $0.23^{+0.36}_{-0.04}$ | $3.5^{+1.6}_{-1.0}$ | $1.7^{+2.5}_{-0.2}$ | $1.9^{+0.9}_{-0.6}$ | $1.1^{+1.7}_{-0.1}$ | $5.3 \pm 1.6$ |
| J0037−0942 | $0.74^{+0.17}_{-0.28}$ | $0.74^{+0.10}_{-0.43}$ | $0.43^{+0.09}_{-0.17}$ | $0.48^{+0.08}_{-0.31}$ | $4.7^{+1.0}_{-1.7}$ | $4.7^{+0.5}_{-2.7}$ | $2.7^{+0.5}_{-1.0}$ | $3.0^{+0.4}_{-1.9}$ | $5.6 \pm 1.7$ |
| J0044+0113 | $0.55^{+0.27}_{-0.15}$ | $0.70^{+0.12}_{-0.46}$ | $0.36^{+0.13}_{-0.14}$ | $0.46^{+0.09}_{-0.31}$ | $3.5^{+1.7}_{-0.9}$ | $4.5^{+0.6}_{-2.9}$ | $2.3^{+0.8}_{-0.9}$ | $2.9^{+0.5}_{-2.0}$ | $6.4 \pm 1.9$ |
| J0109+1500 | $1.08^{+0.29}_{-0.22}$ | $0.79^{+0.45}_{-0.26}$ | $0.64^{+0.14}_{-0.16}$ | $0.49^{+0.30}_{-0.16}$ | $4.9^{+1.2}_{-0.9}$ | $3.5^{+2.0}_{-1.1}$ | $2.9^{+0.6}_{-0.7}$ | $2.2^{+1.3}_{-0.7}$ | $4.7 \pm 1.4$ |
| J0157−0056 | $1.21^{+0.20}_{-0.52}$ | $0.70^{+0.38}_{-0.22}$ | $0.70^{+0.10}_{-0.32}$ | $0.44^{+0.25}_{-0.14}$ | $4.6^{+0.6}_{-1.9}$ | $2.7^{+1.4}_{-0.8}$ | $2.7^{+0.3}_{-1.2}$ | $1.7^{+0.9}_{-0.5}$ | $3.2 \pm 1.0$ |
| J0216−0813 | $0.71^{+0.19}_{-0.28}$ | $0.53^{+0.19}_{-0.15}$ | $0.41^{+0.08}_{-0.14}$ | $0.35^{+0.12}_{-0.11}$ | $4.8^{+1.2}_{-1.8}$ | $3.5^{+1.2}_{-1.0}$ | $2.8^{+0.4}_{-0.9}$ | $2.4^{+0.8}_{-0.7}$ | $4.4 \pm 1.3$ |
| J0252+0039 | $0.52^{+0.07}_{-0.24}$ | $0.43^{+0.08}_{-0.29}$ | $0.29^{+0.06}_{-0.13}$ | $0.26^{+0.05}_{-0.17}$ | $4.5^{+0.5}_{-2.0}$ | $3.6^{+0.6}_{-2.4}$ | $2.4^{+0.4}_{-1.1}$ | $2.2^{+0.4}_{-1.5}$ | $4.8 \pm 1.4$ |
| J0330−0020 | $0.99^{+0.11}_{-0.27}$ | $0.80^{+0.10}_{-0.28}$ | $0.55^{+0.07}_{-0.13}$ | $0.51^{+0.05}_{-0.16}$ | $4.3^{+0.2}_{-1.1}$ | $3.4^{+0.3}_{-1.1}$ | $2.4^{+0.2}_{-0.5}$ | $2.2^{+0.1}_{-0.7}$ | $4.3 \pm 1.3$ |
| J0405−0455 | $0.73^{+0.43}_{-0.23}$ | $0.90^{+0.32}_{-0.57}$ | $0.44^{+0.24}_{-0.17}$ | $0.58^{+0.24}_{-0.36}$ | $4.0^{+2.4}_{-1.2}$ | $5.0^{+1.7}_{-3.2}$ | $2.4^{+1.3}_{-0.9}$ | $3.3^{+1.3}_{-2.0}$ | $6.9 \pm 2.0$ |
| J0728+3835 | $0.50^{+0.25}_{-0.08}$ | $0.46^{+0.30}_{-0.15}$ | $0.28^{+0.16}_{-0.03}$ | $0.34^{+0.16}_{-0.16}$ | $3.2^{+1.5}_{-0.4}$ | $2.9^{+1.9}_{-0.9}$ | $1.8^{+1.0}_{-0.1}$ | $2.2^{+1.0}_{-1.0}$ | $5.5 \pm 1.6$ |
| J0737+3216 | $0.77^{+0.09}_{-0.17}$ | $0.62^{+0.09}_{-0.19}$ | $0.44^{+0.04}_{-0.11}$ | $0.42^{+0.05}_{-0.16}$ | $5.4^{+0.3}_{-1.1}$ | $4.4^{+0.4}_{-1.2}$ | $3.1^{+0.1}_{-0.7}$ | $2.9^{+0.2}_{-1.1}$ | $4.5 \pm 1.4$ |
| J0822+2652 | $0.93^{+0.10}_{-0.16}$ | $0.82^{+0.09}_{-0.33}$ | $0.53^{+0.06}_{-0.11}$ | $0.51^{+0.08}_{-0.23}$ | $5.7^{+0.2}_{-0.8}$ | $5.0^{+0.3}_{-2.0}$ | $3.3^{+0.1}_{-0.6}$ | $3.1^{+0.4}_{-1.4}$ | $5.2 \pm 1.5$ |
| J0903+4116 | $0.87^{+0.09}_{-0.31}$ | $0.73^{+0.09}_{-0.29}$ | $0.49^{+0.06}_{-0.21}$ | $0.46^{+0.06}_{-0.21}$ | $4.9^{+0.2}_{-1.7}$ | $4.1^{+0.3}_{-1.6}$ | $2.8^{+0.2}_{-1.1}$ | $2.5^{+0.2}_{-1.2}$ | $3.7 \pm 1.2$ |
| J0912+0029 | $0.51^{+0.11}_{-0.09}$ | $0.54^{+0.16}_{-0.22}$ | $0.30^{+0.06}_{-0.06}$ | $0.34^{+0.12}_{-0.15}$ | $5.9^{+1.1}_{-0.9}$ | $6.2^{+1.7}_{-2.5}$ | $3.5^{+0.7}_{-0.7}$ | $3.9^{+1.3}_{-1.7}$ | $5.9 \pm 1.7$ |
| J0935−0003 | $0.52^{+0.09}_{-0.12}$ | $0.37^{+0.12}_{-0.10}$ | $0.30^{+0.04}_{-0.08}$ | $0.24^{+0.07}_{-0.07}$ | $5.1^{+0.7}_{-1.1}$ | $3.6^{+1.1}_{-0.9}$ | $2.9^{+0.2}_{-0.7}$ | $2.4^{+0.7}_{-0.6}$ | $4.3 \pm 1.3$ |
| J0936+0913 | $0.71^{+0.21}_{-0.17}$ | $0.58^{+0.47}_{-0.25}$ | $0.40^{+0.12}_{-0.10}$ | $0.35^{+0.31}_{-0.14}$ | $3.6^{+1.0}_{-0.8}$ | $2.9^{+2.4}_{-1.2}$ | $2.0^{+0.6}_{-0.5}$ | $1.8^{+1.6}_{-0.7}$ | $5.6 \pm 1.7$ |
| J0946+1006 | $0.51^{+0.06}_{-0.11}$ | $0.44^{+0.06}_{-0.25}$ | $0.30^{+0.03}_{-0.08}$ | $0.29^{+0.04}_{-0.17}$ | $5.3^{+0.3}_{-1.0}$ | $4.5^{+0.4}_{-2.6}$ | $3.1^{+0.1}_{-0.8}$ | $3.0^{+0.3}_{-1.7}$ | $5.3 \pm 1.6$ |
| J0956+5100 | $0.74^{+0.11}_{-0.33}$ | $0.65^{+0.13}_{-0.42}$ | $0.42^{+0.06}_{-0.19}$ | $0.41^{+0.07}_{-0.27}$ | $5.2^{+0.6}_{-2.3}$ | $4.6^{+0.8}_{-2.9}$ | $3.0^{+0.3}_{-1.3}$ | $2.9^{+0.4}_{-1.9}$ | $5.2 \pm 1.5$ |
| J0959+4416 | $0.82^{+0.19}_{-0.20}$ | $0.83^{+0.09}_{-0.51}$ | $0.52^{+0.06}_{-0.14}$ | $0.50^{+0.07}_{-0.30}$ | $4.5^{+0.9}_{-1.0}$ | $4.6^{+0.1}_{-2.8}$ | $2.9^{+0.1}_{-0.7}$ | $2.8^{+0.3}_{-1.6}$ | $5.2 \pm 1.6$ |
| J0959+0410 | $0.65^{+0.13}_{-0.15}$ | $0.65^{+0.23}_{-0.31}$ | $0.38^{+0.10}_{-0.11}$ | $0.42^{+0.13}_{-0.21}$ | $6.6^{+1.1}_{-1.5}$ | $6.6^{+2.2}_{-3.1}$ | $3.9^{+1.0}_{-1.1}$ | $4.2^{+1.3}_{-2.1}$ | $6.3 \pm 1.9$ |
| J1016+3859 | $0.63^{+0.16}_{-0.18}$ | $0.49^{+0.35}_{-0.13}$ | $0.37^{+0.09}_{-0.09}$ | $0.35^{+0.20}_{-0.11}$ | $4.4^{+1.0}_{-1.2}$ | $3.4^{+2.4}_{-0.8}$ | $2.6^{+0.6}_{-0.6}$ | $2.4^{+1.4}_{-0.7}$ | $5.9 \pm 1.7$ |
| J1020+1122 | $0.46^{+0.19}_{-0.12}$ | $0.37^{+0.21}_{-0.08}$ | $0.30^{+0.08}_{-0.10}$ | $0.22^{+0.14}_{-0.04}$ | $3.8^{+1.5}_{-0.9}$ | $3.1^{+1.8}_{-0.6}$ | $2.5^{+0.6}_{-0.8}$ | $1.8^{+1.1}_{-0.3}$ | $4.8 \pm 1.4$ |
| J1023+4230 | $0.71^{+0.09}_{-0.15}$ | $0.66^{+0.08}_{-0.20}$ | $0.41^{+0.05}_{-0.09}$ | $0.43^{+0.05}_{-0.16}$ | $6.1^{+0.5}_{-1.1}$ | $5.6^{+0.3}_{-1.6}$ | $3.5^{+0.2}_{-0.6}$ | $3.7^{+0.2}_{-1.3}$ | $5.6 \pm 1.7$ |
| J1029+0420 | $0.81^{+0.32}_{-0.27}$ | $1.00^{+0.21}_{-0.53}$ | $0.47^{+0.22}_{-0.16}$ | $0.63^{+0.14}_{-0.34}$ | $4.2^{+1.6}_{-1.3}$ | $5.2^{+1.0}_{-2.7}$ | $2.4^{+1.1}_{-0.8}$ | $3.3^{+0.7}_{-1.8}$ | $6.5 \pm 1.9$ |
| J1100+5329 | $0.49^{+0.32}_{-0.15}$ | $0.26^{+0.37}_{-0.06}$ | $0.34^{+0.12}_{-0.15}$ | $0.17^{+0.24}_{-0.04}$ | $3.1^{+2.0}_{-0.9}$ | $1.7^{+2.4}_{-0.3}$ | $2.2^{+0.8}_{-0.9}$ | $1.1^{+1.6}_{-0.2}$ | $4.5 \pm 1.4$ |
| J1106+5228 | $1.01^{+0.38}_{-0.29}$ | $0.92^{+0.42}_{-0.29}$ | $0.60^{+0.19}_{-0.16}$ | $0.64^{+0.28}_{-0.23}$ | $5.1^{+1.9}_{-1.4}$ | $4.7^{+2.1}_{-1.4}$ | $3.0^{+0.9}_{-0.8}$ | $3.3^{+1.4}_{-1.1}$ | $6.7 \pm 2.0$ |
| J1112+0826 | $0.70^{+0.10}_{-0.10}$ | $0.66^{+0.17}_{-0.17}$ | $0.40^{+0.06}_{-0.05}$ | $0.43^{+0.11}_{-0.12}$ | $6.1^{+0.7}_{-0.7}$ | $5.7^{+1.4}_{-1.4}$ | $3.5^{+0.4}_{-0.3}$ | $3.8^{+0.9}_{-1.0}$ | $4.9 \pm 1.5$ |
| J1134+6027 | $0.65^{+0.28}_{-0.25}$ | $0.74^{+0.20}_{-0.48}$ | $0.41^{+0.12}_{-0.18}$ | $0.46^{+0.20}_{-0.31}$ | $4.6^{+1.9}_{-1.7}$ | $5.2^{+1.4}_{-3.3}$ | $2.9^{+0.8}_{-1.2}$ | $3.2^{+1.3}_{-2.1}$ | $6.0 \pm 1.8$ |
| J1142+1001 | $0.59^{+0.24}_{-0.13}$ | $0.49^{+0.30}_{-0.11}$ | $0.38^{+0.12}_{-0.12}$ | $0.34^{+0.18}_{-0.09}$ | $3.6^{+1.5}_{-0.7}$ | $3.0^{+1.8}_{-0.6}$ | $2.3^{+0.7}_{-0.7}$ | $2.1^{+1.1}_{-0.5}$ | $5.3 \pm 1.6$ |
| J1143−0144 | $0.46^{+0.09}_{-0.06}$ | $0.50^{+0.07}_{-0.17}$ | $0.28^{+0.05}_{-0.04}$ | $0.33^{+0.06}_{-0.12}$ | $5.8^{+1.0}_{-0.5}$ | $6.3^{+0.6}_{-2.1}$ | $3.5^{+0.5}_{-0.3}$ | $4.1^{+0.6}_{-1.4}$ | $6.5 \pm 1.9$ |
| J1153+4612 | $0.51^{+0.34}_{-0.07}$ | $0.33^{+0.56}_{-0.06}$ | $0.28^{+0.23}_{-0.03}$ | $0.21^{+0.38}_{-0.04}$ | $2.8^{+1.8}_{-0.2}$ | $1.8^{+3.0}_{-0.3}$ | $1.5^{+1.2}_{-0.1}$ | $1.1^{+2.1}_{-0.2}$ | $5.7 \pm 1.7$ |
| J1204+0358 | $0.43^{+0.14}_{-0.12}$ | $0.38^{+0.25}_{-0.09}$ | $0.26^{+0.07}_{-0.06}$ | $0.24^{+0.16}_{-0.06}$ | $3.7^{+1.2}_{-0.9}$ | $3.2^{+2.1}_{-0.7}$ | $2.2^{+0.5}_{-0.5}$ | $2.1^{+1.4}_{-0.5}$ | $5.9 \pm 1.7$ |
| J1205+4910 | $0.61^{+0.19}_{-0.19}$ | $0.64^{+0.20}_{-0.38}$ | $0.35^{+0.10}_{-0.11}$ | $0.40^{+0.16}_{-0.25}$ | $5.1^{+1.5}_{-1.5}$ | $5.4^{+1.7}_{-3.2}$ | $3.0^{+0.8}_{-0.9}$ | $3.4^{+1.3}_{-2.1}$ | $5.4 \pm 1.6$ |
| J1213+6708 | $0.71^{+0.21}_{-0.18}$ | $0.75^{+0.23}_{-0.31}$ | $0.42^{+0.11}_{-0.09}$ | $0.46^{+0.19}_{-0.19}$ | $5.2^{+1.5}_{-1.2}$ | $5.5^{+1.7}_{-2.3}$ | $3.1^{+0.8}_{-0.6}$ | $3.4^{+1.4}_{-1.4}$ | $6.3 \pm 1.9$ |
| J1218+0830 | $0.64^{+0.21}_{-0.14}$ | $0.73^{+0.16}_{-0.32}$ | $0.39^{+0.12}_{-0.08}$ | $0.48^{+0.12}_{-0.23}$ | $4.9^{+1.5}_{-1.0}$ | $5.6^{+1.1}_{-2.4}$ | $2.9^{+0.8}_{-0.6}$ | $3.7^{+0.9}_{-1.7}$ | $6.2 \pm 1.8$ |
| J1250+0523 | $1.04^{+0.27}_{-0.33}$ | $1.00^{+0.13}_{-0.63}$ | $0.60^{+0.12}_{-0.16}$ | $0.58^{+0.16}_{-0.34}$ | $4.1^{+1.0}_{-1.2}$ | $4.0^{+0.3}_{-2.5}$ | $2.4^{+0.4}_{-0.6}$ | $2.3^{+0.6}_{-1.3}$ | $5.2 \pm 1.6$ |
| J1402+6321 | $0.73^{+0.13}_{-0.12}$ | $0.72^{+0.26}_{-0.29}$ | $0.42^{+0.09}_{-0.07}$ | $0.47^{+0.16}_{-0.21}$ | $6.0^{+0.9}_{-0.8}$ | $5.9^{+2.0}_{-2.3}$ | $3.4^{+0.7}_{-0.5}$ | $3.9^{+1.3}_{-1.7}$ | $5.5 \pm 1.6$ |
| J1403+0006 | $0.84^{+0.27}_{-0.30}$ | $0.88^{+0.17}_{-0.54}$ | $0.51^{+0.12}_{-0.21}$ | $0.51^{+0.15}_{-0.30}$ | $4.1^{+1.2}_{-1.4}$ | $4.3^{+0.7}_{-2.6}$ | $2.5^{+0.5}_{-1.0}$ | $2.5^{+0.7}_{-1.4}$ | $5.6 \pm 1.7$ |
| J1416+5136 | $0.75^{+0.09}_{-0.17}$ | $0.62^{+0.08}_{-0.21}$ | $0.43^{+0.04}_{-0.10}$ | $0.41^{+0.05}_{-0.14}$ | $5.2^{+0.3}_{-1.0}$ | $4.3^{+0.4}_{-1.4}$ | $3.0^{+0.1}_{-0.6}$ | $2.9^{+0.2}_{-0.9}$ | $4.7 \pm 1.4$ |
| J1420+6019 | $1.01^{+0.43}_{-0.36}$ | $0.84^{+0.60}_{-0.24}$ | $0.63^{+0.23}_{-0.18}$ | $0.55^{+0.38}_{-0.16}$ | $4.5^{+1.9}_{-1.5}$ | $3.7^{+2.6}_{-1.0}$ | $2.8^{+1.0}_{-0.8}$ | $2.4^{+1.7}_{-0.7}$ | $7.0 \pm 2.1$ |
| J1430+4105 | $0.38^{+0.07}_{-0.13}$ | $0.26^{+0.11}_{-0.15}$ | $0.21^{+0.03}_{-0.08}$ | $0.19^{+0.04}_{-0.12}$ | $4.0^{+0.6}_{-1.3}$ | $2.8^{+1.1}_{-1.6}$ | $2.3^{+0.3}_{-0.8}$ | $2.1^{+0.4}_{-1.3}$ | $4.8 \pm 1.4$ |
| J1436−0000 | $0.77^{+0.24}_{-0.30}$ | $0.35^{+0.53}_{-0.07}$ | $0.45^{+0.12}_{-0.19}$ | $0.22^{+0.33}_{-0.05}$ | $4.4^{+1.3}_{-1.7}$ | $2.0^{+3.0}_{-0.3}$ | $2.6^{+0.6}_{-1.0}$ | $1.3^{+1.9}_{-0.2}$ | $4.8 \pm 1.4$ |
| J1443+0304 | $1.00^{+0.32}_{-0.44}$ | $0.92^{+0.54}_{-0.42}$ | $0.57^{+0.18}_{-0.20}$ | $0.60^{+0.40}_{-0.28}$ | $5.4^{+1.7}_{-2.3}$ | $5.0^{+2.9}_{-2.2}$ | $3.0^{+1.0}_{-1.0}$ | $3.3^{+2.1}_{-1.5}$ | $6.2 \pm 1.8$ |
| J1451−0239 | $0.97^{+0.22}_{-0.48}$ | $0.94^{+0.41}_{-0.48}$ | $0.60^{+0.09}_{-0.32}$ | $0.64^{+0.28}_{-0.35}$ | $5.6^{+1.2}_{-2.7}$ | $5.4^{+2.3}_{-2.7}$ | $3.5^{+0.4}_{-1.9}$ | $3.7^{+1.6}_{-2.0}$ | $6.3 \pm 1.9$ |
| J1525+3327 | $0.68^{+0.09}_{-0.21}$ | $0.53^{+0.07}_{-0.18}$ | $0.36^{+0.05}_{-0.10}$ | $0.35^{+0.04}_{-0.10}$ | $5.0^{+0.5}_{-1.5}$ | $3.9^{+0.4}_{-1.3}$ | $2.7^{+0.3}_{-0.7}$ | $2.6^{+0.1}_{-0.7}$ | $4.2 \pm 1.3$ |
| J1531−0105 | $0.70^{+0.15}_{-0.14}$ | $0.70^{+0.23}_{-0.24}$ | $0.42^{+0.08}_{-0.08}$ | $0.44^{+0.17}_{-0.15}$ | $4.7^{+0.9}_{-0.8}$ | $4.7^{+1.5}_{-1.6}$ | $2.8^{+0.5}_{-0.5}$ | $2.9^{+1.1}_{-1.0}$ | $5.9 \pm 1.8$ |
| J1538+5817 | $0.84^{+0.08}_{-0.19}$ | $0.71^{+0.11}_{-0.26}$ | $0.46^{+0.05}_{-0.07}$ | $0.46^{+0.06}_{-0.18}$ | $5.0^{+0.1}_{-1.0}$ | $4.2^{+0.5}_{-1.5}$ | $2.7^{+0.0}_{-0.3}$ | $2.7^{+0.3}_{-1.0}$ | $6.1 \pm 1.8$ |
| J1621+3931 | $0.75^{+0.12}_{-0.25}$ | $0.66^{+0.10}_{-0.21}$ | $0.42^{+0.07}_{-0.13}$ | $0.46^{+0.05}_{-0.19}$ | $5.1^{+0.7}_{-1.6}$ | $4.5^{+0.5}_{-1.4}$ | $2.9^{+0.3}_{-0.8}$ | $3.1^{+0.1}_{-1.3}$ | $5.1 \pm 1.5$ |
| J1627−0053 | $0.61^{+0.12}_{-0.13}$ | $0.46^{+0.30}_{-0.12}$ | $0.34^{+0.07}_{-0.06}$ | $0.29^{+0.20}_{-0.09}$ | $4.4^{+0.7}_{-0.8}$ | $3.3^{+2.1}_{-0.8}$ | $2.5^{+0.5}_{-0.4}$ | $2.1^{+1.4}_{-0.6}$ | $5.5 \pm 1.6$ |
| J1630+4520 | $0.69^{+0.07}_{-0.10}$ | $0.61^{+0.07}_{-0.14}$ | $0.39^{+0.04}_{-0.05}$ | $0.39^{+0.04}_{-0.11}$ | $6.1^{+0.3}_{-0.7}$ | $5.3^{+0.3}_{-1.1}$ | $3.4^{+0.1}_{-0.3}$ | $3.4^{+0.2}_{-0.9}$ | $5.1 \pm 1.5$ |
| J1636+4707 | $0.59^{+0.18}_{-0.12}$ | $0.46^{+0.26}_{-0.10}$ | $0.35^{+0.09}_{-0.09}$ | $0.28^{+0.22}_{-0.05}$ | $3.5^{+1.0}_{-0.7}$ | $2.7^{+1.6}_{-0.5}$ | $2.1^{+0.5}_{-0.5}$ | $1.7^{+1.3}_{-0.3}$ | $5.3 \pm 1.6$ |
| J2238−0754 | $0.64^{+0.25}_{-0.25}$ | $0.72^{+0.23}_{-0.36}$ | $0.37^{+0.14}_{-0.14}$ | $0.48^{+0.17}_{-0.25}$ | $4.7^{+1.8}_{-1.8}$ | $5.4^{+1.6}_{-2.6}$ | $2.8^{+1.0}_{-1.0}$ | $3.6^{+1.2}_{-1.9}$ | $6.2 \pm 1.8$ |
| J2300+0022 | $0.60^{+0.07}_{-0.11}$ | $0.49^{+0.09}_{-0.19}$ | $0.32^{+0.04}_{-0.06}$ | $0.32^{+0.05}_{-0.14}$ | $5.8^{+0.3}_{-0.9}$ | $4.8^{+0.8}_{-1.8}$ | $3.1^{+0.3}_{-0.5}$ | $3.1^{+0.4}_{-1.3}$ | $5.3 \pm 1.6$ |
| J2303+1422 | $0.63^{+0.13}_{-0.09}$ | $0.68^{+0.12}_{-0.22}$ | $0.37^{+0.07}_{-0.05}$ | $0.44^{+0.10}_{-0.16}$ | $5.7^{+1.1}_{-0.5}$ | $6.1^{+0.9}_{-1.9}$ | $3.3^{+0.5}_{-0.3}$ | $4.0^{+0.9}_{-1.4}$ | $6.0 \pm 1.8$ |
| J2321−0939 | $0.90^{+0.26}_{-0.18}$ | $0.96^{+0.26}_{-0.36}$ | $0.53^{+0.20}_{-0.10}$ | $0.64^{+0.19}_{-0.23}$ | $6.6^{+1.8}_{-1.2}$ | $7.1^{+1.8}_{-2.6}$ | $3.9^{+1.5}_{-0.7}$ | $4.7^{+1.3}_{-1.6}$ | $6.8 \pm 2.0$ |
| J2341+0000 | $0.73^{+0.13}_{-0.28}$ | $0.64^{+0.25}_{-0.43}$ | $0.41^{+0.08}_{-0.19}$ | $0.43^{+0.17}_{-0.30}$ | $5.9^{+0.9}_{-2.2}$ | $5.1^{+1.9}_{-3.4}$ | $3.3^{+0.6}_{-1.5}$ | $3.4^{+1.3}_{-2.4}$ | $5.7 \pm 1.7$ |

strong-lensing modeling of 22 galaxies of the SLACS survey. Both these previous results are consistent, within the uncertainties, with our average values, if a Salpeter IMF is adopted (see Table 2). We note that a sample of early-type lens galaxies more distant than the galaxies considered in this study would be invaluable in probing galactic regions beyond one effective radius. Future wide-field imaging surveys [e.g., Panoramic Survey Telescope & Rapid Response System (Pan-STARRS) and Large Synoptic Survey Telescope (LSST)] will provide samples of this kind. By modeling these data, we should be able to observe, at these large radii, a statistically significant increase in the dark matter fraction, as already suggested by several analyses of high-redshift lenses (e.g., Treu & Koopmans 2004; Grillo et al. 2008c).

Starting from previous considerations, we decide to quantify the amount and the distribution of dark matter in the sample of lenses. Thus, we model the cumulative mass profile of the luminous component as a de Vaucouleurs profile [see Eq. (1)] and the cumulative mass profile of the dark component as a power law [$M_{\rm D}(\leq R) \propto R^{\eta}$] or a pseudo-Jaffe profile $\left[M_{\rm D}(\leq R) \propto \left(\sqrt{R^2 + R_1^2} - \sqrt{R^2 + R_2^2}\right)\right]$. The pseudo-Jaffe pro-

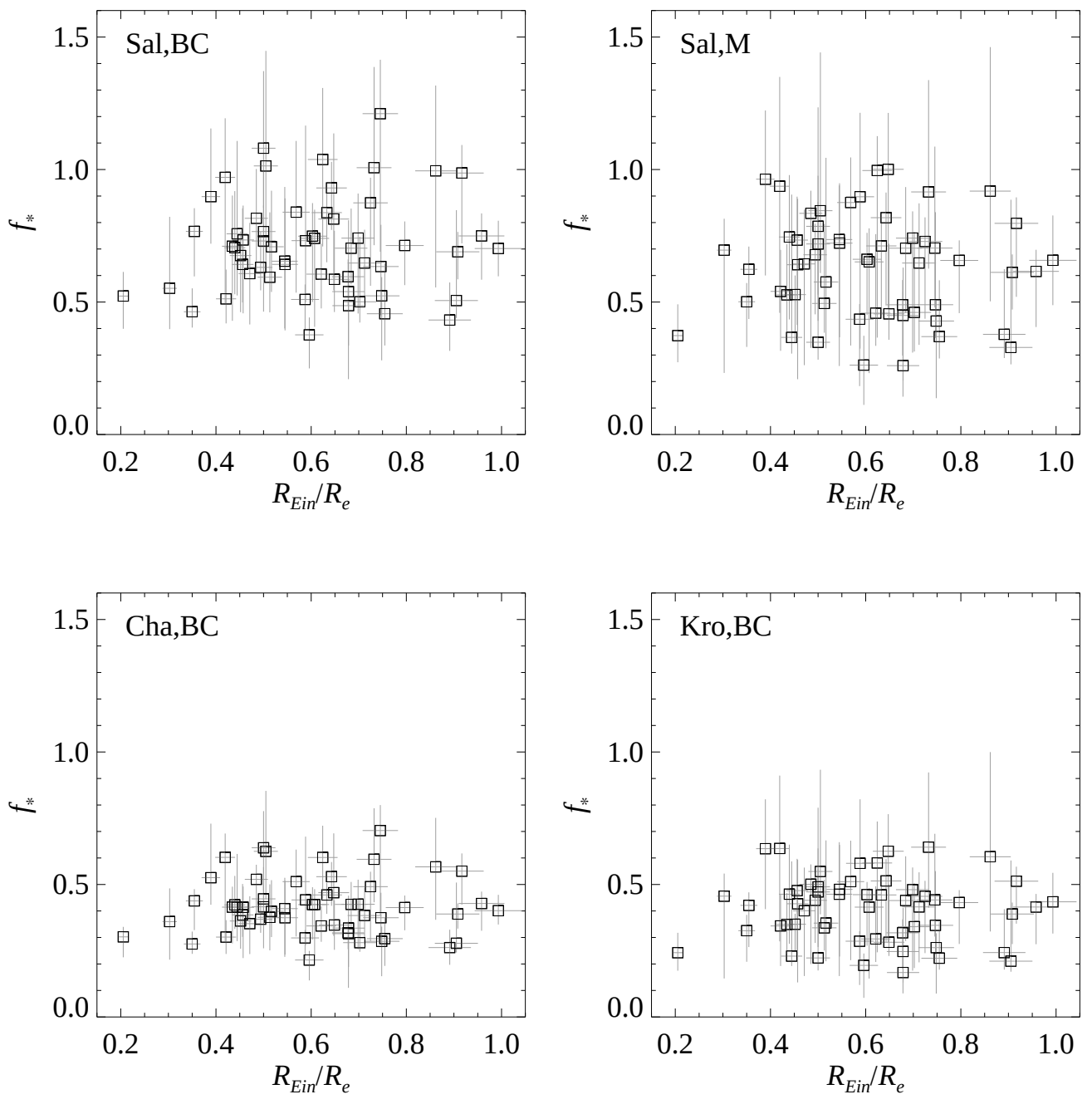

**Fig. 10.** The fraction of mass in the form of stars ($f_*$) enclosed within the disk defined by the Einstein radius (with 1 $\sigma$ error bars) versus the Einstein radius (in units of the effective radius) of the 57 early-type grade-A lens galaxies of the SLACS survey. The photometric mass estimates are obtained by different composite stellar-population models [Bruzual & Charlot (BC) and Maraston (M)] and IMFs [Salpeter (Sal), Chabrier (Cha), Kroupa (Kro)].

**Table 6.** The 1 $\sigma$ confidence intervals of the mass in the form of dark matter.

| | Sal,BC/Sal,M | Cha,BC/Kro,M |
|---|---|---|
| $M_D[(R_{Ein}/R_e) \leq 0.2]\,(M_*)$ | [0.04, 0.22] | [0.18, 0.43] |
| $M_D[(R_{Ein}/R_e) \leq 0.6]\,(M_*)$ | [0.19, 0.31] | [0.58, 0.70] |
| $M_D[(R_{Ein}/R_e) \leq 1.0]\,(M_*)$ | [0.20, 0.41] | [0.63, 1.12] |

Notes – In each aperture, the lower and upper limits of the intervals are chosen as the smallest and the largest values predicted, at 1 $\sigma$ confidence level, by the best-fit *power-law* and *pseudo-Jaffe* models. These values are given in units of the luminous mass, that is determined by adopting a Salpeter (*left*) or Chabrier/Kroupa (*right*) IMF.

file derives its name from the three-dimensional density distribution [$\rho(r) \propto (r^2 + r_1^2)^{-1}(r^2 + r_2^2)^{-1}$], which is similar to that of the Jaffe (1983) model [$\rho(r) \propto r^{-2}(r + r_2)^{-2}$]. We optimize the parameters of the dark matter profiles to reproduce the observed fraction of mass in the form of stars and estimate the errors in the best-fit profiles by means of 1000 Monte Carlo simulations.

The results are shown in Table 6 and in Figs. 11, 12, 13, and 14. We find that the best-fit values of the exponent $\eta$ of the power-law cumulative mass profile range between 0.32 and 0.54, suggesting that in the inner regions, the dark matter density profile may be significantly steeper than a NFW profile

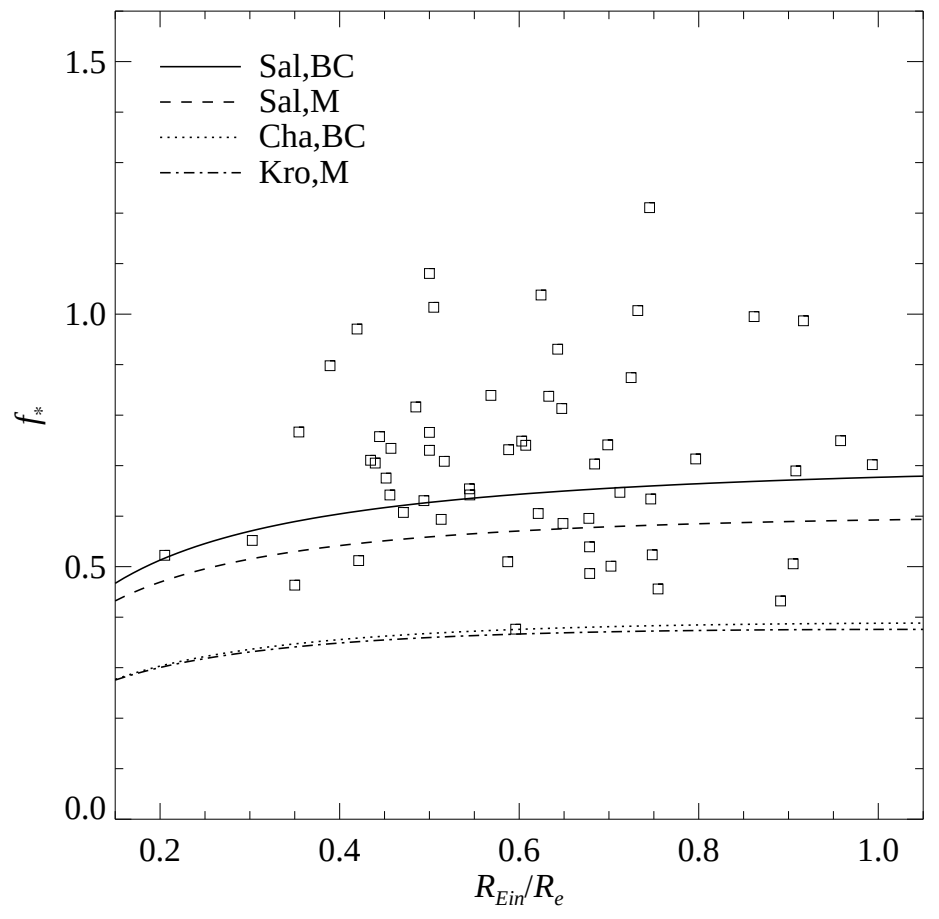

**Fig. 11.** The fraction of mass in the form of stars ($f_*$) enclosed within the disk defined by the Einstein radius versus the Einstein radius (in units of the effective radius) of the 57 early-type grade-A lens galaxies of the SLACS survey as obtained by using Bruzual & Charlot models and a Salpeter IMF. The lines show the best-fit *power-law* models for different composite stellar-population models [Bruzual & Charlot (BC) and Maraston (M)] and IMFs [Salpeter (Sal), Chabrier (Cha), Kroupa (Kro)].

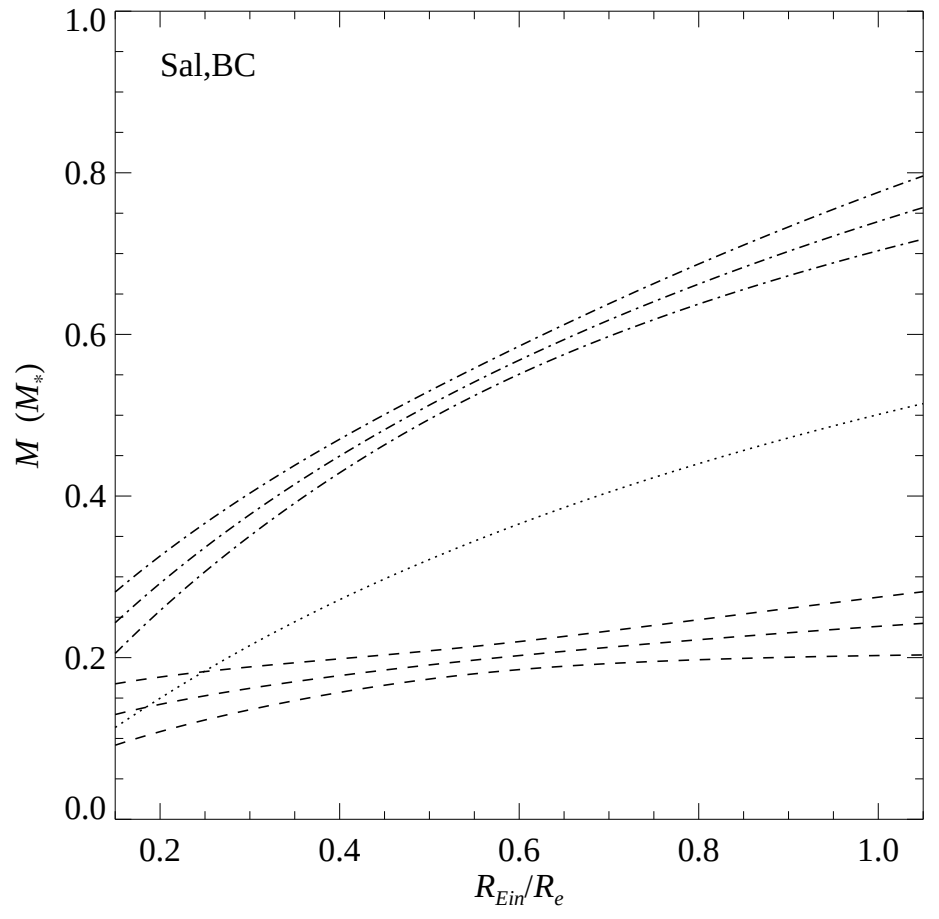

**Fig. 12.** The mass decomposition. The profile of the luminous component is modeled by a de Vaucouleurs profile and is represented by a dotted line; the profile of the dark component is modeled by a power law [$M_D(\leq R) \propto R^{\eta}$] and is represented by a dashed line (with 1 $\sigma$ error bars); the total mass (i.e., the sum of the luminous and dark components) is represented by a dashed-dotted line (with 1 $\sigma$ error bars). The masses are expressed in units of the photometric mass, as obtained by using Bruzual & Charlot models and a Salpeter IMF.

(Navarro et al. 1997). This explains, *a posteriori*, our second choice of a pseudo-Jaffe instead of a NFW profile for the dark matter component. By assuming a pseudo-Jaffe profile, we note that the best-fit values of the core radius $R_1$ are smaller than 0.005. This strengthens our statement about the possible concentration of dark matter in the center of the lens galaxies. The plots confirm the qualitative discussions about the amount of dark matter made by analyzing Fig. 9. In addition, we note that the cumulative total mass profile predicted by all the models is consistent within the errors with a function increasing linearly with $R$ (i.e., with what is expected for a $1/r^2$ "isothermal" total density distribution), independent of the assumed profile of the dark matter component. As already mentioned, the fraction of dark matter required to reproduce the total mass measurements by assuming either a Chabrier or Kroupa IMF is significantly higher than for a Salpeter IMF. In particular, in the former case, the mass in the form of dark matter within $R_{\rm e}$ can be more than the estimated mass in the form of stars, and, in the latter case, the dark mass within $R_{\rm e}$ is less than about 40% of the luminous mass (see Table 6).

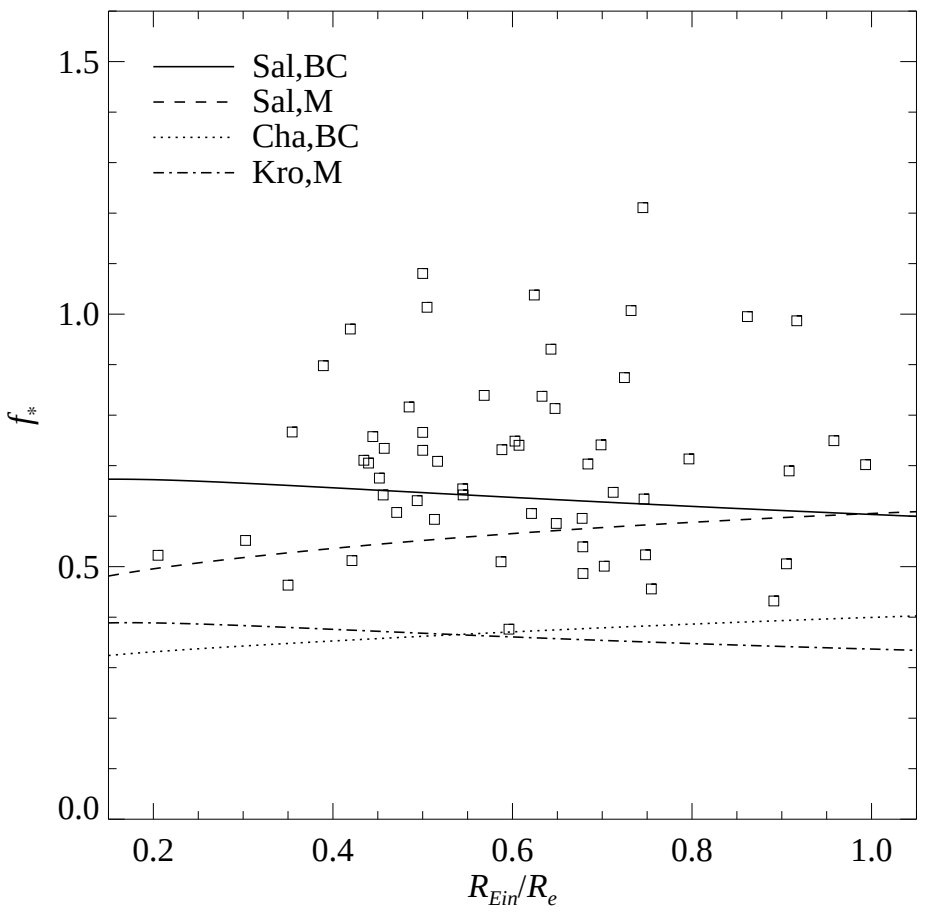

**Fig. 13.** The fraction of mass in the form of stars ($f_*$) enclosed within the disk defined by the Einstein radius versus the Einstein radius (in units of the effective radius) of the 57 early-type grade-A lens galaxies of the SLACS survey as obtained by using Bruzual & Charlot models and a Salpeter IMF. The lines show the best-fit *pseudo-Jaffe* models for different composite stellar-population models [Bruzual & Charlot (BC) and Maraston (M)] and IMFs [Salpeter (Sal), Chabrier (Cha), Kroupa (Kro)].

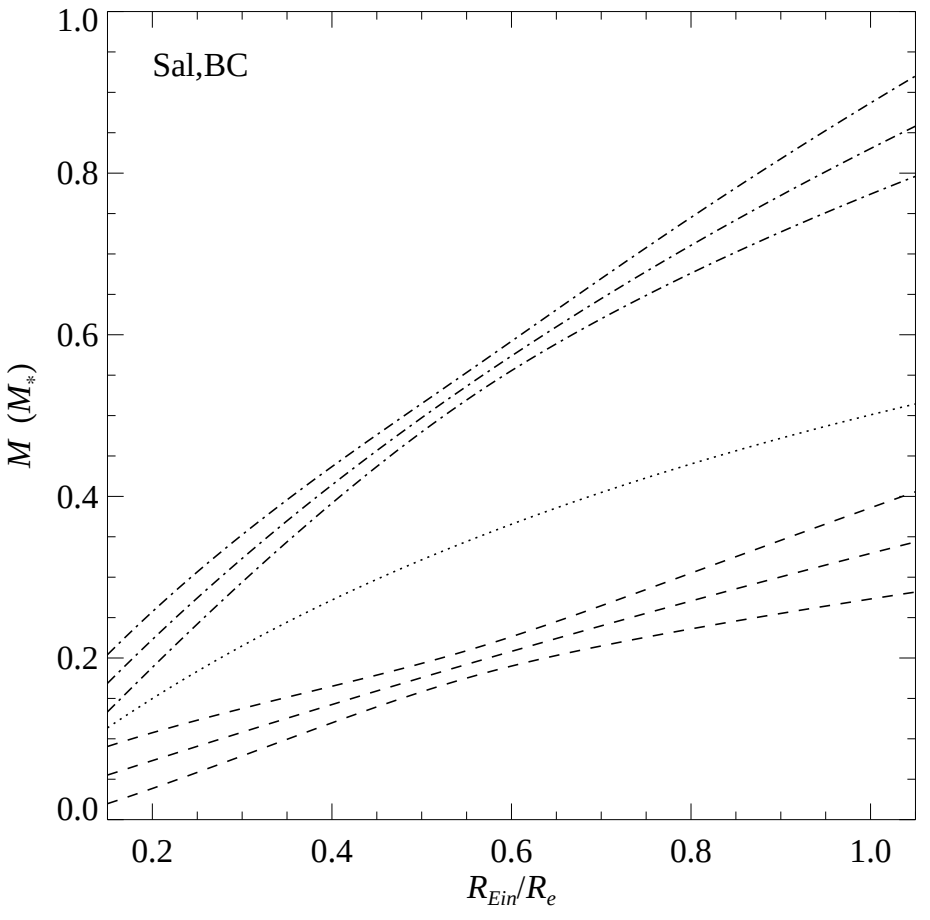

**Fig. 14.** The mass decomposition. The profile of the luminous component is modeled by a de Vaucouleurs profile and is represented by a dotted line; the profile of the dark component is modeled by a function obtained by integrating a pseudo-Jaffe profile $\left[M_{\rm D}(\leq R) \propto \left(\sqrt{R^2+R_1^2} - \sqrt{R^2+R_2^2}\right)\right]$ and is represented by a dashed line (with 1 $\sigma$ error bars); the total mass (i.e., the sum of the luminous and dark components) is represented by a dashed-dotted line (with 1 $\sigma$ error bars). The masses are expressed in units of the photometric mass, as obtained by using Bruzual & Charlot models and a Salpeter IMF.

In Grillo et al. (2008a), we demonstrated that inside the Einstein radii of early-type lens galaxies, a Salpeter IMF implies photometric estimates of mass in the form of stars that agree with estimates from lensing+dynamics. We also found that the photometric masses obtained by assuming either a Kroupa or Chabrier IMF are probably underestimated (based on the hypothesis of solar metallicity). The extremely large amount of dark matter predicted here by choosing either a Kroupa or Chabrier IMF seems to suggest the same conclusion.

This test, in addition to that already conducted and shown in Fig. 8 for the mass-to-light ratios, implies that photometric mass measurements obtained by adopting a Salpeter IMF are more reliable. Thus, we conclude that the dark-matter estimates of the first two models must be considered more realistic than those of the last two models.

## 5. Summary and conclusions

We have studied the photometric mass properties and the luminous and total mass content of the 57 grade-A early-type lens galaxies of the SLACS survey. First, we have modeled the galaxy SEDs, obtained from the SDSS multicolor photometry, to measure the mass in the form of stars present in the lenses. Then, with the aim of differentiating between the luminous and dark matter components inside the Einstein radii of the lensing systems, we have combined our results with the best-fit values of the lens models developed by Bolton et al. (2008a).

The sample galaxies are located at redshifts of between 0.06 and 0.51 and the values of the Einstein angle of the lensing systems vary between 0.69″and 1.78″. The average value of the Einstein radii is equal to $4.1 \pm 0.2$ kpc and this is approximately 0.6 times the average value of the effective radii of the lens galaxies.

In detail, the main results of our study can be summarized as follows:

- The distribution of the lens galaxies on the size-stellar mass and surface stellar mass density-stellar mass planes is consistent with that of massive early-type galaxies in the local Universe.
- The values of the photometric mass of the lens galaxies correlate with the values of the overdensity estimator. This indicates that massive lens galaxies reside preferentially in regions where the density of neighboring galaxies is high, as is indeed true for early-type galaxies in general.
- If the composite stellar-population models are followed in their passive evolution until redshift zero, at this redshift, the predicted values of the $B$-band stellar mass-to-light ratio of the lenses scale with the values of their photometric mass in the same way as the corresponding effective quantities of local early-type galaxies scale according to the FP.

This implies that the dynamical (or total) masses of the lens galaxies are linearly proportional to their respective photometric masses.
- For the lenses, the evolutionary rate value of the rest-frame *B*-band stellar mass-to-light ratio is consistent, within the uncertainties, with that obtained from the FP on the effective mass-to-light ratio. This offers some more evidence of the linear relation between the lens total and photometric masses.
- The values of the rest-frame *B*-band stellar mass-to-light ratio of the lenses, measured from composite stellar-population models with a Salpeter IMF, are consistent, within the errors, with the values derived from the FP, whereas the values obtained by adopting a Chabrier or Kroupa IMF are systematically lower than those inferred by the FP.
- Inside the Einstein radii of the lensing systems, the values of the total projected mass measured from the lens models are linearly proportional to the values of the luminous mass estimated from the SED fits.
- The average values of the ratio of dark to total projected mass inside the Einstein radii of the lensing systems vary approximately from 30% to 60%, depending on the adopted IMF.
- By assuming that the sample galaxies are homologous systems, the value of the projected mass in the form of dark matter enclosed within one effective radius is between 0.2 and 0.4 times the value of the photometric mass obtained with a Salpeter IMF.

According to the first set of results, we emphasize that early-type lens galaxies do not differ from massive early-type non-lens galaxies in terms of the luminous mass properties. This allows us to generalize the final set of results, that refer to the dark mass component and can be obtained only by combining photometric and lensing measurements in lens galaxies, to the overall population of massive early-type galaxies.

*Acknowledgements.*

We thank G. Bertin and S. Seitz for useful comments on this manuscript. We are grateful to L. Koopmans and T. Treu for stimulating conversations.

In this work we have made extensive use of the database provided by the Sloan Digital Sky Survey. Funding for the SDSS and SDSS-II has been provided by the Alfred P. Sloan Foundation, the Participating Institutions, the National Science Foundation, the U.S. Department of Energy, the National Aeronautics and Space Administration, the Japanese Monbukagakusho, the Max Planck Society, and the Higher Education Funding Council for England. The SDSS Web Site is http://www.sdss.org/. The SDSS is managed by the Astrophysical Research Consortium for the Participating Institutions. The Participating Institutions are the American Museum of Natural History, Astrophysical Institute Potsdam, University of Basel, University of Cambridge, Case Western Reserve University, University of Chicago, Drexel University, Fermilab, the Institute for Advanced Study, the Japan Participation Group, Johns Hopkins University, the Joint Institute for Nuclear Astrophysics, the Kavli Institute for Particle Astrophysics and Cosmology, the Korean Scientist Group, the Chinese Academy of Sciences (LAMOST), Los Alamos National Laboratory, the Max-Planck-Institute for Astronomy (MPIA), the Max-Planck-Institute for Astrophysics (MPA), New Mexico State University, Ohio State University, University of Pittsburgh, University of Portsmouth, Princeton University, the United States Naval Observatory, and the University of Washington.